# Drivers of periodicity in population dynamic models of long-lived, large mammals

Marron Clifford McConnell, [1,*] William F. Fagan [1]

[1]Department of Biology, University of Maryland, College Park, MD, USA

[*]Corresponding author: mecmcc@umd.edu

Manuscript type: Article

## Abstract

Population cycles are important components of many natural systems. Most studied in short-lived and small-bodied species, cycles frequently appear to be driven by density-dependent feedbacks. However, compelling evidence of cycles—often more qualitative than quantitative—also exists in large mammals. Among ungulates, both density-dependent vital rates and 'cohort effects' (lasting impacts of birth conditions on fecundity and survival) exist, but the implications of such feedbacks for oscillatory population dynamics have not been explored. Here, we present a synthetic model of ungulate population dynamics, parameterized for barren-ground caribou (*Rangifer tarandus groenlandicus*) and motivated by extensive Indigenous knowledge suggesting decades-long fluctuations in abundance. Caribou herds are theorized to be subject to both cohort effects and density dependence, and we linked these endogenous factors with environmental stochasticity to understand cycling. Using wavelet analysis, we characterized periodic phenomena and performed sensitivity analyses to clarify the drivers and characteristics of population cycles. We found that cohort effects, predominantly those impacting survival, can produce long-period oscillatory behavior across a wide range of environments and demographic structures. Our modeling framework is generalizable to other long-lived, large-bodied species with complex demography, and collectively, these efforts broaden the scope of inquiry into proximal drivers of population cycling.

# Introduction

Much attention has been paid to the dynamics of cycle-like fluctuations in abundance of wild populations driven by diverse mechanisms such as periodic environments (Hunter and Price 1998; Tzuk et al. 2019), tightly coupled consumer-resource interactions (Stenseth et al. 1999), and overcompensatory density dependence (Goodsman et al. 2017). Time delays are often critical to the existence of cycles as the survival or reproduction of the population is temporally decoupled from the immediate conditions (Turchin 1990; Kendall et al. 1999; Beckerman et al. 2002).

Focus has predominantly been on cycles in insect or small mammal systems (Kendall et al. 1999) and it has long been suggested that long-lived species are unlikely to cycle because they have longer generation times with lower birth rates and therefore steadier growth (Herrando-Pérez et al. 2012; Sæther et al. 2013). Indeed, for each of the factors thought to facilitate cycling in some small-bodied species, arguments have been put forward explaining why that factor is unlikely to engender periodic population dynamics in larger-bodied species. For example, large herbivores occupy and move across larger spatial areas, diluting the strength of the coupling between resource and consumer (Ofstad et al. 2016; Pichon et al. 2024). Longer generation times can reduce the pace of consumer-resource feedbacks (Pichon et al. 2024) in species that may already have looser relationships with vegetation and predation (Hopcraft et al. 2012). Hypotheses like the Jarman-Bell principle explain why body condition may be less sensitive to fluctuations in quality of forage (Müller et al. 2013). Finally, allometric scaling of predator functional response parameters (Coblentz et al. 2025) may clarify why tight population feedbacks with predators are less common for large animals.

Obvious exceptions to these arguments exist, such as for herbivore populations strongly linked to predator populations (Christianson and Creel 2014; Chamaillé-Jammes et al. 2019).

Nevertheless, ecologists are used to thinking that large-bodied species only cycle due to strong seasonal forcing (Forchhammer and Boertmann 1993), rare tight feedback across trophic levels (Post et al. 2002), or colonization of new habitat (Caughley 1970). Such studies may consider density dependence lagged by a couple of years, but few examine density dependence at a lag of a decade or beyond.

One barrier to empirical study is feasibility of data collection. Because cycle periods scale with generation time or body size (Peterson et al. 1984), it might take several decades to collect even one time series of appropriate length. Given the relative dearth of such studies, less attention has been paid to long-period population cycles despite evidence of their existence (Grenfell et al. 1992; Larter and Nagy 2001; Gunn 2003; Zagorodniuk 2022). Many long-lived species are subject to mechanisms known to cause cycling in short-lived small-bodied species (Clutton-Brock et al. 1992; Gaillard et al. 1997; Reynolds 1998) and documentation of quasi-cyclic dynamics exist for ungulate species not commonly understood to undergo population cycles, including white-tailed deer (*Odocoileus virginianus*, Fryxell et al. 1991), moose (*Alces alces*, Peterson et al. 1984; Post et al. 2002), and elk (*Cervus canadensis*, Starns et al. 2014).

Cohort effects, the lasting impacts of conditions experienced at birth and during early development on lifelong individual fitness (Lindström 1999), represent a potential source of time delays that could entrain cycling. Early-life conditions leading to cohort effects could include environmental states, density, predator abundance, or a combination of factors influencing life history traits. For example, calves born in 'good' conditions (i.e., born to mothers with favorable body condition, at lower densities, amid advantageous environmental conditions, or other such factors) have, on average, increased survival and/or reproduction in red deer (*Cervus elaphus*, Mysterud et al. 2002; Loison et al. 2004), roe deer (*Capreolus capreolus*, Gaillard et al. 2003; Douhard et al. 2014; Plard et al. 2015), bighorn sheep (*Ovis canadensis*, Pigeon et al. 2017), Norwegian reindeer (*Rangifer tarandus tarandus*, Weladji and Holand 2003), and Svalbard reindeer (*Rangifer tarandus platyrhynchus*, Douhard et al. 2016; Pigeon et al. 2019). This approach differs from so-called cohort cycles (e.g. see de Roos et al., 2003), in which population dynamics are dominated by alternating life stages with cycle periods related to the length of delay in reproductive maturity. In contrast, cohort effects are driven not by dynamic maturation delays but by dynamic lifetime fitness resulting from differential early life conditions.

When transient conditions have lasting impacts on individuals' lifetime fitness in long-lived species, delays in the population-level response to changing conditions can result. To see this, suppose conditions are poor for a stretch, and cohorts born in those years will eventually have lower adult survival. As these three cohorts age, fewer individuals than usual would survive, reducing overall cohort calf production. This reduction would not take effect for years, when those cohorts reach reproductive maturity, and could last until those cohorts die out over a decade later. Environmental and density trends can therefore accumulate through successive cohorts, a long-term feedback process only possible for long-lived species. Especially when considered in context of effects of age and immediate conditions, cohort effects and their interactions can drive a population away from equilibrial dynamics and even entrain cycles (Kaitala et al. 1996; Grøtan et al. 2009; Betini et al. 2017).

Here, we demonstrate that common assumptions about long-lived species—weak coupling across trophic levels, greater capacity to buffer against fluctuations in environmental quality, long-life span with innately lower growth rates—are not incompatible with population cycles but simply incompatible with short-period cycles. We highlight that if there is the potential for vital rates in the present to be impacted by density or environmental conditions in the distant past, then it is precisely greater longevity of individuals that opens up the potential for population cycles. Our goal here is to clarify which hypotheses of cohort effects and density dependence, in concert with what demographic structures and under what kinds of environments, are most likely to entrain long-period population cycles.

## *A Case Study in Caribou*

Caribou (*Rangifer tarandus*), known in Eurasia as wild reindeer, are a keystone species (COSEWIC 2016) found globally across the entire Arctic region. The migratory barren-ground caribou (*R. t. groenlandicus*) herds of Alaska, mainland Canada, and Greenland shape ecosystems through herbivory and nutrient cycling (van der Wal 2006; Gunn et al. 2011; Rickbeil et al. 2015), landscape alteration (Zalatan et al. 2006), and as prey for Arctic predators (Bergerud 1996; Gau et al. 2002). Their seasonal movements represent the largest terrestrial migration worldwide (Joly et al. 2019), elevating the impact of this subspecies across massive spatial scales.

Barren-ground caribou are considered to have oscillatory population dynamics on multi-decadal timescales (Griffith et al. 2002; Gunn 2003; Zalatan et al. 2006; Łutsël K'é Dene First Nation 2023). Most North American subpopulations are showing significant decreases in abundance after highs in the 1980s and 1990s, with few showing signs of recovery (Fauchald et al. 2017). However, limited data availability impedes statistical assessment of how current crashes relate to purported 'natural' population cycles. For example, less than half of the herds in North America have sufficient numbers of population estimates to fit a wavelet function, and low sample sizes reduce statistical power when fitting sinusoidal cyclic models to population estimates (Bongelli et al. 2020).

Without sufficient data to empirically explore cycling in more depth, mechanistic modeling offers a path to understanding long-period cycles under varying hypotheses of the complex relationships between environment and demography. Such modeling permits consideration of the relative importance of diverse factors within a single context. Endogenous drivers of oscillations in caribou are expected to include density dependence (Tyler 2010; Tveraa et al. 2013) and cohort effects (Douhard et al. 2016; Pigeon et al. 2019). Exogenous drivers are expected to operate primarily through accumulated body conditions (Cameron et al. 1993; Parker et al. 2009; Mallory et al. 2018). Temperature and precipitation are most obviously associated with body condition through their impact on vegetation (Albon et al. 2017; Veiberg et al. 2017) but can also have non-vegetation-mediated effects on body condition. For example, deeper snow can lengthen the duration of migratory journeys and increase energy expenditure en route (Duquette 1988; Gurarie et al. 2019), whereas higher temperatures in longer summers worsen insect harassment with severe consequences (Witter et al. 2012; Mallory and Boyce 2018).

To understand how this litany of effects combine to influence population dynamics, we modeled vital rates as nonlinear functions representing different hypotheses of cohort effects and density dependence as they relate to environmental drivers. We then assessed these various models for their propensity to yield population cycles to clarify the conditions under which long-period cycles can be expected.

# Methods

### *Summary*

Our approach synthesizes deterministic density-dependent structured matrix models (e.g. Caswell, 1989) with stochastic but unstructured models (e.g., Roughgarden, 1975). We opted to add structure back to unstructured models that link stochasticity to density dependence through a time-dependent carrying capacity; our focus was on transient, non-equilibrial dynamics.

We created a discrete-time, age-structured, female-only demographic model considering survival and fecundity as a function of age and environmental or density conditions at birth and during adulthood. We explored different hypotheses by changing the characteristics of the timeseries representing the environment (the environmental "scenario") and the underlying functions linking environment to vital rates. We created a factorial design whereby each triplet combination of hypotheses concerning density dependence, cohort effects, and environmental scenario was evaluated. Each such three-part combination was termed a "model," and collectively we simulated population abundance timeseries (population trajectories) across 88 different models (see SI Table A-2).

## *Population Dynamics Equations and baseline rates*

We divided the population into stages traditionally considered by caribou biologists (Boulanger and Gunn 2007): calves $(C)$, yearlings $(Y)$, and adults $(A)$. Stages were further divided by age ($\alpha$), with single year classes for calves (zero-year-olds) and yearlings (one-year-olds) and adults comprising ages two through the maximum age of 16 (Cuyler and Østergaard 2005; Lee et al. 2015). As caribou are a birth-pulse species with an annual synchronized calving period (Nagy and Campbell 2012; COSEWIC 2016), we selected a discrete-time post-birth model where the year begins immediately post-calving. $A(\alpha, t)$, the number of adults of age $\alpha$ in year $t$, is the product of the number of individuals of age $\alpha - 1$ in year $t - 1$ times their survival rate:

$$A(\alpha, t) = \begin{cases} s(\alpha - 1, t - 1)Y(t - 1), & \alpha = 2 \\ s(\alpha - 1, t - 1)A(\alpha - 1, t - 1), & \alpha > 2 \end{cases} \quad (1)$$

where $s(\alpha, t)$ denotes survival that can be age- and time-specific. The total number of adults is the sum across all adult age classes,

$$A(t) = \sum_{\alpha \geq 2} A(\alpha, t)\,. \quad (2)$$

The number of yearlings in year $t$ is the product of the number of calves in year $t-1$ and their survival rate,

$$Y(t) = s(\alpha = 0, t-1)C(t-1). \tag{3}$$

The number of female calves born in year $t$ to individuals of age $\alpha$ is the product of the number of individuals of age $\alpha$ in year $t$ and their fecundity that year divided by two (assuming a 1:1 sex ratio at birth),

$$C(t) = \frac{1}{2}\sum_{\alpha \geq 2} f(\alpha, t)A(\alpha, t) \tag{4}$$

where $f(\alpha, t)$ denotes fecundity that can be age- and time-specific. The shift in time indexing ($t-1$ vs $t$) is a consequence of the post-birth-pulse model formulation. Multiple births and twinning are negligible in caribou (McDonald and Martell 1981; COSEWIC 2016), red deer (Guinness et al. 1978), elk (Hudson and Haigh 2002) and bighorn sheep (Festa- Bianchet et al. 2019). Caribou also have a delayed age of first reproduction, with nearly all females not calving until age three.

These equations can also be expressed in matrix form where the transition matrix, $\boldsymbol{L}$, is a function of $t$ and $\alpha$, though the latter only appears implicitly in the matrix (each entry is function of pairs of years corresponding to the current year and the birth year for each age-class). The matrix equation can be summarized as

$$\boldsymbol{N(t)} = \boldsymbol{L(t, t-1)N(t-1)}, \tag{5}$$

where

$$\boldsymbol{N(t)} = \begin{bmatrix} C(t) \\ Y(t) \\ A(2,t) \\ \vdots \\ A(15,t) \\ A(16,t) \end{bmatrix} \tag{6a}$$

and

$$\boldsymbol{L(t,t-1)} = \begin{bmatrix} 0 & \frac{1}{2}s(1,t-1)f(2,t) & \frac{1}{2}s(2,t-1)f(3,t) & \dots & \frac{1}{2}s(15,t-1)f(16,t) & 0 \\ s(0,t-1) & 0 & 0 & \dots & 0 & 0 \\ 0 & s(1,t-1) & 0 & \dots & 0 & 0 \\ 0 & 0 & s(2,t-1) & \dots & 0 & 0 \\ 0 & 0 & 0 & \ddots & 0 & 0 \\ 0 & 0 & 0 & \dots & s(15,t-1) & 0 \end{bmatrix}. (6b)$$

We selected baseline rates based on publicly available data from the Bathurst herd (Adamczewski et al. 2021) and Beverly herd (Thomas and Barry 1990), and adjusted them such that without environmental forcing, cohort effects, and density dependence, the population neither grows nor shrinks. We explored different hypotheses by changing functional forms governing vital rates, leading to realized age-specific rates that differed from baselines.

Realized vital rates were the product of a core vital rate function $\mathrm{G}(\cdot)$ (see *Cohort Effect Hypotheses* below) and a modifier $g(\alpha)$ introduced to ensure ecological realism,

$$f(\alpha,t) = G^f(\cdot)g^f(\alpha) \quad (7\text{a})$$

$$s(\alpha,t) = G^s(\cdot)g^s(\alpha) \quad (7\text{b})$$

The modifier $g^f(\alpha)$ ensured that fecundity of non-adults was always zero and fecundity of two-year olds was below that of other adults,

$$g^f(\alpha) = \begin{cases} 0, & \alpha < 2 \\ 0.5, & \alpha = 2 \\ 1, & \alpha > 2 \end{cases}. \quad (8)$$

The modifier $g^s(\alpha)$ ensured that survival of non-adults was below that of adults and survival of the terminal age class was fixed to 0,

$$g^S(\alpha) = \begin{cases} 0.8, & \alpha \le 2 \\ 1, & 2 < \alpha < 16 \\ 0, & \alpha = 16 \end{cases}. \quad (9)$$

As discussed in *Cohort Effect Hypotheses*, we explored hypotheses regarding cohort effects by changing the functional form of $G(\cdot)$, incorporating different combinations of conditions at birth, and in the current year. As discussed in *Environmental Scenarios*, we explored hypotheses regarding environmental scenarios by changing the timeseries of environmental values, $\varepsilon(t)$. Finally, as discussed in *Density Dependence Hypotheses*, we explored hypotheses regarding the lack or presence of density dependence by making conditions at birth, $\zeta_{\text{birth}}(\alpha, t)$, and adulthood, $\zeta_{\text{curr}}(t)$ be functions solely of the environment or of density with an environmentally determined carrying capacity.

## *Cohort Effect Hypotheses*

Drawing on the work of Pigeon et al. (2019) and Engqvist and Reinhold (2016), we explored four different hypotheses, each corresponding to a functional form governing the translation of external conditions into vital rates. These hypotheses are not specific to any one species and can apply to wide variety of taxa. First, the "current conditions" (CC) hypothesis can be considered the base model without cohort effects in that only the time-local conditions matter. Second, Grafen (1988) referred to the "silver spoon" (SS) effect in which individuals born under good conditions enjoy higher vital rates than their counterparts born under poor conditions regardless of adult conditions (also see Lindström 1999). Third, the "environmental saturation" (ES) hypothesis, theorized by Engqvist and Reinhold (2016), posits that conditions in adulthood dampen the cohort effect such that differences in birth conditions are negligible at the extremes of adult conditions. Pigeon et al. (2019) found support for the environmental saturation hypothesis for fecundity in Svalbard reindeer. Last, the "environmental matching" (EM) hypothesis, also known as the "predictive adaptive response" (PAR) hypothesis, was proposed by Gluckman and Hanson (2004*a*, 2004*b*) in the context of human development. This hypothesis posits that conditions are not innately

'good' or 'bad', rather individuals do best in adult conditions that match the conditions of their birth. The underlying rationale has been criticized in humans (Wells 2007; Hayward and Lummaa 2013) and is said to lack empirical grounding (Uller et al. 2013; Douhard et al. 2014); however, it has been documented in at least one system (the soil mite *Sancassania berlesei*, Beckerman et al. 2003).

Mathematically, we denote current 'conditions' in any given year as $\zeta_{\text{curr}}(t)$, and birth 'conditions' as $\zeta_{\text{birth}}(\alpha, t)$ for individuals aged $\alpha$ in year $t$. The actual $\zeta(\cdot)$ functions were altered according to different hypotheses of density dependence, detailed in *Density Dependence Hypotheses*.

We further extended the CC, SS, ES and EM hypotheses by combining them with Age structure, resulting in four new hybrid hypotheses: CCA, SSA, ESA and EMA. Each hypothesis had a corresponding function $G(\alpha, t)$, detailed in Table 1, which was substituted into the vital rate functions (equations 7a and 7b). Biological interpretation of each hypotheses functions can be found in SI Appendix A.

As shown in Table 1, parameters $\{\beta_i^j\}, j \in \{f, s\}, i \in \{0,1,2,3,4,5\}$ controlled the relative importance of each term in determining the realized fecundity and survival, respectively. To avoid conflating birth effects and current effects for young individuals, no terms with birth effects were included in the calculation of rates for 2-year-olds and younger.

The exact shape of the functional form of the cohort effects depends on both the parameters and the $\zeta(\cdot)$ functions, which presented an issue for the ESA cohort effect in models with density dependence. To circumvent this, the ESA hypothesis with its explicit interaction terms was only considered for the environment-only models and for the density-dependent models, an ESA-like form can be recovered with large values for parameters corresponding to the SSA hypothesis. For details, see SI Appendix A.

### *Environmental Scenarios*

We created demographic matrices to simulate population trajectories by linking the density dependence and cohort effect hypotheses with timeseries of environmental drivers. Each simulated environment timeseries, $\varepsilon(t)$, was rescaled such that $\varepsilon(t) \in [-1,1]$. We fit auto-regressive integrated moving average (ARIMA) models to a variety of empirical timeseries of environmental values and then simulated timeseries from the ARIMA fits.

We selected a mix of broad climate and local weather indices known to influence caribou dynamics at different scales (Boulanger and Adamczewski 2017). The final set of environmental drivers we selected were: the Summer, Winter, and Whole-year averages of Arctic Oscillation (AO) index, the Summer Oestrid Index (OI, a composite metric of temperature and wind as a proxy for the severity of insect harassment during the summer, Russell et al. 2013), June 20$^{th}$ growing-degree-days (GDD), March 31$^{st}$ snow depth and Winter freezing rain. Detailed explanations of each scenario can be found in SI Appendix B. We also tested model behavior under a constant scenario of zeros to better contextualize the stochastic, empirically motivated scenarios. Greater discussion of this constant environment scenario can be found in SI Appendix C.

AO indices were retrieved from the National Oceanic and Atmospheric Administration (NOAA/CPC 2026). Other environmental metrics were provided by the Modern Era Retrospective Analysis for Research and Applications (MERRA) database maintained by the CircumArctic Rangifer Monitoring Association (CARMA, Russell et al. 2013).

### *Density Dependence Hypotheses*

Density dependence can present on different timescales and with different compensatory strengths. The cohort effect hypotheses provide a pathway for linking 'conditions' to vital rates, allowing the consideration of circumstances where density is a 'condition' itself. We developed four hypotheses of how density relates to both birth conditions, $\zeta_{\text{birth}}(\alpha, t)$, and current conditions, $\zeta_{\text{curr}}(t)$: (i) the absence of density dependence, (ii) delayed density dependence, in which only birth conditions are influenced by density, (iii) immediate density dependence, in which only current conditions are influenced by density and (iv) full (both delayed and immediate) density dependence, in which both birth and current conditions are influenced by density. Each hypothesis corresponded to different forms of $\zeta_{\text{birth}}(\alpha, t)$ and $\zeta_{\text{curr}}(t)$.

We first considered the absence of density dependence, in which both $\zeta(\cdot)$ terms were functions of environmental values unmediated by density. For this environment-only case, birth effects and current effects were simply the environmental values corresponding to, respectively, the birth year and the current year,

$$\zeta_{\text{birth}}(\alpha, t) = \varepsilon(t - \alpha)\,. \tag{10a}$$

$$\zeta_{\text{curr}}(t) = \varepsilon(t)\,. \tag{10b}$$

For models with density dependence, we opted for an explicit carrying capacity based approach, as in the Ricker model (1954), due to its more concrete interpretability. Carrying capacities can fluctuate with changes in habitat conditions. Consequently, building from a stochastic modeling framework introduced by Roughgarden (1975), we introduced a dynamic carrying capacity $K(t)$ as function of $\varepsilon(t)$, the simulated timeseries of environmental values,

$$K(t) = 2K_0 \frac{e^{\varepsilon(t)}}{1 + e^{\varepsilon(t)}} \tag{11}$$

where $K_0$ is a scale multiplier that represents carrying capacity of adult females in the absence of environmental drivers; throughout this study, we fixed $K_0 = 150{,}000$. The resulting $K(t)$ was larger than $K_0$ only when $\varepsilon(t) > 0$ and conversely $K(t) < K_0$ only when $\varepsilon(t) < 0$. Because we bounded $\varepsilon(t)$ to be within [-1,1], the realized carrying capacity $K(t)$ is also bounded. This framework posits that environmental drivers act through density dependence; the environment influences population dynamics by altering the carrying capacity each year, essentially mediating the intensity of intraspecific competition (Roughgarden 1975).

All possible equations for realized fecundity and survival are the output of the inverse-logit transform (see Table 1) so additive impacts on the untransformed scale become multiplicative changes with a variable scaling factor on the function's range (the scale of the vital rates). This non-linear multiplicative mapping ensures realized fecundity and survival are guaranteed to stay on the interval [0,1]. Therefore, the modifier term of the logistic map and Ricker model can be used without running into issues – for details, see SI Appendix D.

For models with full density dependence, both $\zeta(\cdot)$ terms in a year are functions of density and environment-dependent carrying capacity. For individuals $\alpha$-years old in year $t$,

$$\zeta_{\text{birth}}(\alpha, t) = 1 - \frac{A(t-\alpha)}{K(t-\alpha)} \; . \tag{12a}$$

$$\zeta_{\text{curr}}(t) = 1 - \frac{A(t)}{K(t)} \; . \tag{12b}$$

The interpretation is that when abundance of adults is low, these $\zeta(\cdot)$ terms are high, resulting in a multiplicative increase on the scale of the vital rate. When adult abundance is at carrying capacity, these $\zeta(\cdot)$ terms are zero, and vital rates stay at baseline. Finally, when adult abundance is above carrying capacity, these $\zeta(\cdot)$ terms are negative, resulting in a multiplicative reduction on the scale of the vital rate.

For models with delayed density dependence, only $\zeta_{\text{birth}}(\alpha, t)$ was a function of density while $\zeta_{\text{curr}}(t)$ was a function of the environment unmediated by density, and vice versa for models with immediate density dependence. These $\zeta(\cdot)$ terms were then substituted into the functions in Table 1, used to calculate realized age-specific vital rates. Note that abundances can exceed $K(t)$ since current density is not the sole determinant of vital rates.

### *Assessing Periodicity: Wavelet Analysis*

To assess periodicity present in our simulated timeseries, we relied upon wavelet analyses (Roesch and Schmidbauer 2018). This approach uncovered periodicity and/or environmental synchrony in population dynamics of a rocky intertidal community of New Zealand (Benincà et al. 2015), breeding success of Antarctic seabirds (Jenouvrier et al. 2005), and abundance of North American porcupines (Klvana et al. 2004), among others.

The wavelet transform, like the windowed Fourier approach, employs a time-local decomposition of a signal (here, the timeseries of population abundance) into components of different periods. Unlike the windowed Fourier approach, the wavelet transform dynamically partitions the signal with rectangular cells whose dimensions shift via changes in a scale parameter, resulting in the optimal time-frequency resolution trade-off (Lau and Weng 1995; Mallat 1999; Cazelles et al. 2008). We considered several wavelets, including the widely used Mexican hat (Ricker 1940) and Morlet wavelets (Morlet et al. 1982), ultimately selecting the Morlet wavelet due to 1) its better handling of noise, 2) its particular time-frequency tradeoff that favors frequency resolution over temporal resolution, and 3) its ability to quantify time-local phase and amplitude (Cazelles et al. 2008).

Prior to wavelet analysis, we trimmed off the first 100 years of each timeseries of adult abundance (to avoid false-positives resulting from initial conditions), then log-transformed each timeseries and standardized the values by subtracting the corresponding means. Furthermore, we excluded from analysis the so-called cone of influence (COI) that quantifies methodologically imposed artifacts (Torrence and Compo 1998). Statistical significance was understood via comparison to sets of 50 bootstrapped timeseries referenced against a first-order autoregressive (AR1) process (Ives et al. 2010) that we found sufficient for generating null distributions. We computed p-values for the null hypothesis that, for any specific frequency, there was no component of that frequency in the original timeseries whose power exceeded that which would be expected had the timeseries been generated by an AR1 process. We used a significance threshold of 0.1. All wavelet analysis was performed in R 4.3.1 using the *WaveletComp* package version 1.1 (Roesch and Schmidbauer 2018).

When one population abundance timeseries from one model (an "individual run") is referred to as being 'periodic,' 'resulting in periodicity' or similar, this means the population timeseries had at least one frequency with statistically significant global wavelet power. We created summary metrics, detailed in Table 2, quantifying the strength and longevity of periodic phenomena relating to individual runs of a model and used maxima and averages of individual run metrics to evaluate periodicity of entire models.

For each run, we assessed the global wavelet power of every frequency for statistical significance and, of those deemed significant, identified the frequency with the highest power. We termed the associated period $\omega$ and associated global wavelet power $\psi$. This value can be thought of as a measure of the overall strength of the dominant component of periodicity for that run. We then determined $\tau$, the longest continuous stretch of years for which the local wavelet power of $\omega$ remained statistically significant.

We used the individual run metrics to summarize the outcomes of entire models. First, we calculated $\rho$ for each model as the proportion of its runs that were periodic. Second, we identified the individual run with the maximum significant global wavelet power and labeled the period, power and longevity metrics of that run with corresponding uppercase Greek letters; that is, we selected the individual run with the highest $\psi$ value across all periodic runs and called this value $\Psi$. We used $\mathrm{T}$ and $\Omega$ to represent, respectively, the longevity and period associated with $\Psi$. These three absolute metrics are best understood as the strongest periodicity possible resulting from a particular model and its corresponding longevity and period.

### *Sensitivity Analysis via Sobol' Indices*

We employed Sobol' sensitivity analyses of our results to identify those factors with the greatest potential to drive long-term periodic dynamics. This variance-based technique (Sobol' 1993; Jansen 1999; Saltelli et al. 1999) has been used in dynamical systems applications including population viability analysis (Ellner and Fieberg 2003). Unlike other variance-based methods such as partial rank correlation coefficients (PRCC), Sobol' sensitivity analysis can be used when the relationship between system inputs and outputs is nonmonotonic, and it can separate first-order effects from higher-order effects for each input parameter (Homma and Saltelli 1996; Saltelli et al. 1999). This method is thus advantageous given the nonlinear interactions among age, birth effects and current effects explored here.

The first-order index quantifies the reduction in the total variance of the output metric that would result were the parameter in question to be fixed. The total-order index jointly quantifies the first-order effect of a parameter together with all its interactions with other parameters. The proportion of the variance attributable to higher-order interactions between parameters can be calculated as 1 minus the sum of all first-order indices.

Sobol' sensitivity analysis was performed in R 4.3.1 with the *Sensobol* package version 1.1.5 (Puy et al. 2022) with standard safeguards involving bootstrapping and checks for asymptotic convergence. Additionally, we used Linear Discriminant Analysis (LDA) to validate qualitative results (see SI Appendix E).

### *Simulating Population Timeseries*

We explored model behavior using a sequence of three rounds of computational modeling, each corresponding to a different set of conditions for generating timeseries. First, we eliminated models that failed to (i) result in periodicity over a majority of parameter space (defined as $\rho \geq 0.5$) and (ii) yield at least one run with 'strong' periodicity (defined as $\Psi \geq 1$). Only the remaining models were carried forward into subsequent rounds. Second, we subjected the reduced set of models to global sensitivity analysis, identified specific parameter sets resulting in the strongest periodicity, and verified that results were robust to different realizations of the environmental scenarios. For the third and final round, we explored how demographic assumptions impacted model outcomes by performing global sensitivity analysis on models with a reduction in the number of age classes and different overall herd fecundity $f_0$ (which resulted in different age-specific baselines $f_0(\alpha)$). Population dynamics were simulated for 1000 years or until the abundance of adult females declined below 1 or exceeded $10^6$.

We performed simulations by fixing representative timeseries for each environmental scenario and varying $\{\beta_i^j\}$, the parameters controlling the relative strength of the various terms in determining fecundity and survival (Table 1). Each model's output consisted of population trajectories simulated from different parameter sets with identical environments. Qualitative outcomes were robust to different example realizations of each environmental scenario, as detailed in SI Appendix B.

The first simulation round was designed to eliminate models with no or weaker tendency to elicit periodicity. For each model, we drew 5000 unique sets of $\{\beta_i^j\}$ from uniform distributions across respective plausible ranges via Latin Hypercube Sampling (Carnell 2022). Seven of 88 models met both criteria—all of them full density dependence SSA/ESA models—and these were advanced into the second and third rounds of investigation.

In the second (sensitivity analysis) round, we employed methods from Saltelli et al. (2010) to draw 30,000 distinct $\{\beta_i^j\}$ from uniform distributions across respective plausible ranges via quasi-random sampling. We then identified parameter sets giving rise to strong periodicity and explored the shapes of those functional forms—for each model, we fixed each parameter at the average value across all individual runs in the 10$^{th}$ percentile for global wavelet power.

In the third round we evaluated sensitivity to assumptions in model structure. We picked a single example model, the Whole-year AO full density dependence SSA/ESA model, from the reduced set of the previous round and constructed five additional models for comparison. The first two additional models kept the maximum age of 16, however we simulated populations using an $f_0 = 0.78$ and $f_0 = 0.66$ instead of the original $f_0 = 0.87$. For the final three additional models, we adjusted the maximum age from 16 to 12 $(\alpha \leq 12)$ and we simulated populations using $f_0 \in \{0.66, 0.78, 0.87\}$. Baseline rates were adjusted to stabilize population growth in the absence of environmental forcing, cohort effects, and density dependence. We simulated trajectories and performed Sobol' sensitivity analyses using the procedure from the second round.

# Results

## *Round 1: Elimination of Models*

Density dependence in some capacity was necessary to induce periodic dynamics, as no environment-only model for any scenario displayed meaningful periodicity. In fact, all had $\rho < 0.01$ and most lacked even a single periodic run. Therefore, all environment-only models were removed from any further analysis and are not displayed on figures. Similarly, density dependent models with a fixed carrying capacity (that is, with no environmental variation) also failed to result in meaningful periodicity. Only 11 runs had sustained, undamped oscillations, out of the 80000 total runs across all density-dependent, constant environment models. Either mechanism acting in isolation was unlikely to yield persistent cycles.

Comparisons of results from all models with at least one kind of density dependence are displayed in Figure 1A. Delayed density dependence models resulted in the strongest periodicity (higher $\Psi$), but over fewer combinations of parameters (lower $\rho$). In contrast, immediate density dependence models resulted in weaker periodicity (lower $\Psi$), but over more combinations of parameters (higher $\rho$). Density dependence in adulthood appears to expand the range of parameters giving rise to periodicity while density dependence in birth effects appears to strengthen the cycles themselves. Models with full density dependence yielded cycle strengths in between that of their delayed and immediate density dependence counterparts but resulted in periodicity across the broadest range of parameters. The inclusion of both kinds of density dependence thus widens the space of possible scenarios under which strong cycles may arise.

The inclusion of cohort effects also had a strong impact on the propensity for periodic population dynamics. Current Conditions with Age-structure (CCA) models, hypothesizing a lack of cohort effects, resulted in the weakest and most fleeting periodicity across the narrowest range of parameter space when compared to different model forms within the same scenario. Without the delay stemming from the birth effect term, even density dependence of the current effect term is only capable of inducing weaker and less continuous periodicity.

Applying the criteria $\rho \geq 0.5$ and $\Psi \geq 1$ eliminated all but seven models from further consideration. Many eliminated models were still capable of exhibiting evident strong or persistent periodicity (Figure 2), but only for a greatly reduced set of parameter combinations (**Figure 1**A). In some cases, population timeseries were 'episodically periodic' with fleeting or discontinuous incidences of high amplitude fluctuations that led to reduced global power (Figure 2, also see SI Appendix C).

The seven models satisfying both criteria were the full density dependence and the SSA/ESA cohort effect for each of the stochastic environmental scenarios. These models were periodic across over 75% of parameter combinations (average $\rho = 0.78$) and had strong, lasting cycles (average $\Psi$ of 1.2 and average $\mathrm{T}$ of 606 years).

### *Round 2: Sensitivity Analysis*

We conducted sensitivity analyses on full density dependence SSA models under the constant scenario and the seven empirically motivated scenarios with respect to summary metrics $\psi$ (strength of periodicity, Figure 3a1) and $\tau$ (longevity of periodicity, Figure 3a2) via Sobol' Indices.

When the output metric was $\psi$, across all scenarios, no Sobol' indices (first- or total-order) for fecundity coefficients were significant, but the total-order indices for all three survival coefficients were significant, as were the first-order indices for $\beta_0^s$ (the $s_0(\alpha)$ coefficient). First-order indices for $\beta_1^s$ (the survival $\zeta_{\text{birth}}$ coefficient) were significant for four scenarios (Whole-year

AO, Summer AO, Summer OI, and June 20th GDD), but not the other three. First-order indices for $\beta_2^s$ (the survival $\zeta_{\text{curr}}$ coefficient) were not significant for any scenario. Fecundity terms had minimal impact on the strength of periodicity both independently and via higher-order interactions; survival terms impacted the strength of periodicity, mostly through $\beta_0^s$, $\beta_1^s$, and higher-order interactions of survival terms with other terms. The average sum of first-order indices was 0.47, meaning the variance in $\psi$ was attributable roughly equally to first-order effects and interactions between parameters.

When the output metric was $\tau$, total-order indices for $\beta_0^f$ (the $f_0(\alpha)$ coefficient) were significant for all seven scenarios, the total-order indices for $\beta_2^f$ (the fecundity $\zeta_{\text{curr}}$ coefficient) were significant for three scenarios (Winter AO, June 20th GDD, and March 31st snow), and the total-order index for $\beta_1^f$ (the fecundity $\zeta_{\text{birth}}$ coefficient) was significant for one scenario (Winter AO). Otherwise, no other indices for fecundity terms were significant. For all scenarios, total-order indices for all three survival coefficients were significant and the first-order indices for $\beta_0^s$ were significant for four scenarios (Whole-year AO, Summer AO, Summer OI, and March 31st snow). No other indices for survival terms were significant. Survival terms were responsible for most of the variance in longevity with some significant (but small effect-size) impacts of fecundity terms. The average sum of first-order indices was 0.16; the overwhelming majority of the variance in $\tau$ was attributable to higher-order interactions between parameters.

Across all models, survival terms dominated in influencing both output metrics of strength and longevity of cycles. This result aligns well with previous modeling work and elasticity analyses indicating adult survival to be a primary driver of population dynamics in ungulates (Gaillard et al. 1998). One explanation for the importance of survival in driving cycling behavior is the relative constraint of fecundity; with no avenue for multiple births per female, whether calf recruitment rates are high enough to overshoot carrying capacity is a question of survival.

We fixed a set of parameters, $p_{avg}$, for each model as follows. Out of the set of 30,000 runs for each model from round 2, we took the distribution of all $\psi$ (strength of periodicity per run), selected only those runs with $\psi$ at or above the 90$^{th}$ percentile, and then calculated the average of each parameter. We found that $p_{avg}$ sets for each model consistently had intermediate forms – individuals born in good conditions always enjoyed higher fitness than those born in poor conditions (in line with the SSA hypothesis), but there was some saturation at extremes (in line with the ESA hypothesis) (**Figure 1**B).

We generated an additional 5,000 environmental timeseries $\varepsilon(t)$ for each scenario to confirm that periodicity arising from the previous rounds for these parameter sets were not simply flukes of the particular realizations of each environmental scenario. With these new $\varepsilon(t)$, we simulated 10,000 population trajectories for each of the seven models using two parameter sets known to yield strong periodicity in previous rounds. All 70,000 runs – across all scenarios and both parameter sets—exhibited substantial periodicity. Given this, it appears likely that such parameter sets can drive periodicity across numerous environmental scenarios.

### *Round 3: Altered Age-structure and Overall Herd Fecundity*

We further considered five additional situations in which we reduced the number of age classes and/or imposed different overall herd fecundity $f_0$ (which resulted in different age-specific baselines $f_0(\alpha)$ and $s_0(\alpha)$). We created five new Whole-year AO full density dependence SSA models, each using one of the new structures for age-specific baseline fecundity. We simulated 30,000 population trajectories for these new models, performed sensitivity analyses, and compared the results to the Whole-year AO full density dependence SSA model with the original age-specific baseline fecundity (**Figure 3**B). Minor discrepancies were evident in the Sobol' indices values themselves, yet the overall trend regarding significance and magnitude of impact persisted: survival terms, particularly higher-order interactions of survival terms with other terms, were the primary determinants of both the power and longevity of population cycles. The few indices for fecundity terms with statistical significance had extremely low effect sizes such that any small influences from fecundity terms were dwarfed by those of survival terms. With the same overarching assumptions (the existence of age-structure, delayed onset of reproductive maturity and negligible twinning), the qualitative interpretation that strong cycles are driven by survival rather than fecundity is robust to changes in demographic parameters.

# Discussion

This study has significant implications at multiple levels; accordingly, we discuss our findings in two distinct sections corresponding to different contexts. First, we consider our results at the focal-species-specific level in *Caribou Cycles and Crashes*. Subsequently, in *Large Herbivore Population Cycles Writ Large*, we address our results at the more general, theoretical level. We review what differentiates our results from classic depictions of short-period population cycles, highlighted in **Figure 4**. We argue that our findings contribute to a better understanding of how, when, and why large-bodied, long-lived species might experience population cycles.

## *Caribou Cycles and Crashes*

Although some simulated populations did show declines of 150,000+ individuals, such high amplitudes were not consistent within a single trajectory, were rare across parameter space, and were usually associated with nadir abundances substantially greater than the current low of the Bathurst herd (Adamczewski et al. 2021). A few simulated trajectories exhibited decreasing phases of a cycle with some alignment to the magnitude and rapidity of the Bathurst decline (SI Appendix F). However, even these timeseries only matched the Bathurst decline initially, entering recovery phases while the Bathurst decline accelerated. Furthermore, the simulated and real timeseries of vital rates substantially differed, particularly for survival.

These findings suggest ongoing declines in wild caribou are unlikely to be solely decreasing phases of 'natural' cycles. Instead, our results lend credence to a 'confluence of declines theory' in which ongoing population crashes result from the convergence of cycle nadirs dovetailing with deleterious anthropogenic change (COSEWIC 2016; Fauchald et al. 2017). In this way, our results highlight the importance of considering population cycles and the hidden nonlinear mechanisms that may drive cycles even when populations are declining in ways that do not appear cyclic.

Our simulated environmental scenarios lacked any explicit accelerated downward trend in environment quality, so perhaps lack of alignment between our simulations and observed declines is because other mechanisms or stronger environmental drivers are at play. Future work should account for this and might also include within-cohort variation as well as the between-cohort variation explored here. Features that could be added are the potential waning of birth effects, alternate modifiers for density dependence, and circumstances in which fecundity and survival depend differently on environmental scenarios or involve different forms of cohort effects.

### *Large Herbivore Population Cycles Writ Large*

We explored the potential for and characteristics of population cycles in a large-bodied, long-lived herbivore by mechanistic modeling of putative determinants of demography and oscillatory dynamics. We found that density dependence and temporal delays in vital rates, introduced as cohort effects, were crucial to the emergence of long-period cycles. However, unlike traditional predator-prey models, in which time delays result from the lag in predator response or vegetation regrowth after herbivory, the delayed density-dependence of our models is fundamentally different in that it is endogenous to the herbivore and arises from herbivore population's own demography and dynamic physiological condition across cohorts. The resulting oscillations do not rely on a tight feedback loop across trophic levels and can arise with only a unidirectional, weak environmental driver and the endogenous factors at one trophic level. Under even weakly stochastic conditions, strong, persistent, long-period oscillations were possible across wide swaths of parameter space and diverse external drivers, with the strength and longevity of periodicity largely determined by second- and higher-order interactions among parameters governing adult survival (Figure 3).

Although environmental variation was necessary for population cycling across most parameter combinations, stochasticity was not a sufficient condition for periodicity without delayed density dependence. Similarly, density dependence in the absence of environmental variation could only produce lasting, undamped oscillations in ~0.01% of cases. Instead, long-period cycles in long-lived species resulted from demographic feedbacks spurred by a stochastic environment. The lack of periodicity in models with only environmental forcing make clear that periodicity was not due to simple synchronization of population dynamics with exogenous oscillations.

Models with cohort effects like the Silver Spoon effect with Age-structure (SSA), with little to no saturation of survivorship under extreme good and bad conditions, resulted in stronger periodicity than models with cohort effects like the Environmental Saturation with Age-structure (ESA), in which survivorship saturated as a function of conditions (**Figure 1**B). Still, models with greater saturation could produce strong cycles lasting hundreds of years uninterrupted (Figure 2 and SI Appendix C). Cohort effects with a combination of delayed and immediate density dependence yielded long-period cycles over the greatest variety of parameter combinations; however, the resulting periodicity was weaker than that which resulted from delayed density dependence alone. The inclusion of immediate density dependence buffered the strength and longevity of the periodicity induced by delayed density dependence. Intuitively, this makes sense; individuals, and the population as a whole, are better able to calibrate to the current conditions, lessening the mismatch between vital rates and environmental conditions.

Even when the drivers of cycles are not strong individually—that is, when environmental stochasticity, density dependence, and cohort effects are all weak— the co-occurrence of all three is capable of inducing cycles. Measuring any one factor in isolation may not be sufficient to explain demographic heterogeneity (Descamps et al. 2008) or population dynamics, and an understanding of how multiple factors converge may be necessary. Within a species, evidence for the existence of mechanisms known to contribute to oscillations should be treated as an invitation to consider the possibility of cycling and investigate for other mechanisms.

Different species may experience such mechanisms in diverse ways. Examples of traits whose values are accrued over time or are strongly impacted by early development (and which therefore might introduce cohort effects into the population) include body condition (Gaillard et al. 2003; Loison et al. 2004; Douhard et al. 2016), allometry or body size (Newman 1989; Festa-Bianchet et al. 2004), rate of development (Pavlovska-Teglia et al. 1995; Crespi and Warne 2013), and onset of reproductive maturity (Telfer et al. 2005). Additionally, complex host-pathogen relationships represent possible avenues for cohort effects to emerge. Especially for density-dependent infection dynamics, higher juvenile susceptibility to infection or reduced virulence with host age (Ducrocq et al. 2013), sublethal infections (Koltz et al. 2022) and buildup of environmental reservoirs of pathogens (Sharp and Pastor 2011; Peacock et al. 2022) may be, or contribute to, drivers of population cycles. Accumulated or early-development-dependent traits are critical and may impact demography in many ungulates (Gulland and Fox 1992; Lankester 2002; Loison et al. 2004). Such populations may be more prone to strong and lasting oscillations in abundance or density, but given long lifespans, decades of longitudinal monitoring would be required to document cycles definitively. In contrast, study of lasting impacts of early life conditions is possible on shorter timescales: Pigeon et al. (2019) were able to quantify cohort effects on fecundity in Svalbard reindeer using just 21 years of data on body mass and reproductive success. Additional focus should be given to mechanisms of population cycling, particularly the relationships among survival, age, accumulated traits, and trends in density or habitat conditions. For long-lived species, populations in decline should be investigated for evidence of cycling which could be converging with consequences of anthropogenic global change to accelerate population crashes.

Our analyses suggest that long-period cycles in large bodied species tend to arise from a complex set of interacting demographic feedbacks (**Figure 4**). Therefore, the conditions favoring these cycles are likely more restrictive than those permitting oscillatory dynamics in short-lived species. Ecologists working with long-lived species often attribute population fluctuations to specific ephemeral factors unrelated to population cycling (e.g. Fragoso et al., 2022). For example, if a species has the intrinsic capacity to cycle, but only experiences a brief spate of conditions favorable to cycling, one may incorrectly ascribe fluctuations to extrinsic conditions or stochastic aberrance. If cohort effects are strong, we may even be looking at the wrong stretch of time, focusing on current conditions that are ultimately not responsible for current dynamics. In stark contrast to the dynamics of short-lived species, cyclical population dynamics in long-lived species are strongly determined by time-delayed, nonlinear feedbacks that might long predate current conditions.

Our modeling demonstrates that a population can initially be non-cyclic, subsequently experience quasi-cyclic fluctuations for a period of time, and then resume non-cyclic behavior without an underlying change in the intrinsic relationships and feedbacks within the population. Without the knowledge of generations of Indigenous peoples, caribou herds monitored since the 1980s would not stand out as cycle-related, yet the herds' propensity to cycle may be a necessary and critical factor in understanding their precipitous declines. Identifying species that have the potential to cycle or quasi-cycle, even if we do not currently see cycles, may help clarify otherwise difficult to explain population phenomena.

## Acknowledgements

We thank Anne Gunn and Eliezer Gurarie for discussions about caribou ecology and Wojciech Czaja for discussions about wavelet analyses. We thank the University of Maryland supercomputing resources (https://hpcc.umd.edu) made available for conducting the research reported in this paper. Funding for this research was provided by NSF- NNA: 212727.

## Statement of Authorship

McConnell: Conceptualization, methods development, simulation and code, model analysis and validation, data visualization, writing – original draft. Fagan: Supervision, funding acquisition, conceptualization, methods refinement, writing – review and editing.

## Data and Code Availability

All simulation code and data are available through the *AmNat* collection on Harvard Dataverse at McConnell (2026), DOI: https://doi.org/10.7910/DVN/7IXSFN.

## References Cited Only in the Online Enhancements

# Tables

Table 1: Vital Rate Functional Forms

| Cohort Effect Hypothesis | Functional form for $\boldsymbol{G^s(\alpha,t)}$ |
|---|---|
| Current Conditions with Age-structure (CCA) | $G^s(\alpha,t) = \dfrac{e^{\beta_0^s \mathrm{logit}(s_0(\alpha)) + \beta_2^s \zeta_{\mathrm{curr}}(t)}}{1 + e^{\beta_0^s \mathrm{logit}(s_0(\alpha)) + \beta_2^s \zeta_{\mathrm{curr}}(t)}}$ |
| Silver Spoon with Age-structure (SSA) | $G^s(\alpha,t) = \dfrac{e^{\beta_0^s \mathrm{logit}(s_0(\alpha)) + \beta_1^s \zeta_{\mathrm{birth}}(\alpha,t) + \beta_2^s \zeta_{\mathrm{curr}}(t)}}{1 + e^{\beta_0^s \mathrm{logit}(s_0(\alpha)) + \beta_1^s \zeta_{\mathrm{birth}}(\alpha,t) + \beta_2^s \zeta_{\mathrm{curr}}(t)}}$ |
| Environmental Matching with Age-structure (EMA) | $G^s(\alpha,t) = \dfrac{e^{\beta_0^s \mathrm{logit}(s_0(\alpha)) + \beta_1^s \zeta_{\mathrm{birth}}(\alpha,t) + \beta_2^s \zeta_{\mathrm{curr}}(t) + \beta_3^s \zeta_{\mathrm{birth}}(\alpha,t)\zeta_{\mathrm{curr}}(t)}}{1 + e^{\beta_0^s \mathrm{logit}(s_0(\alpha)) + \beta_1^s \zeta_{\mathrm{birth}}(\alpha,t) + \beta_2^s \zeta_{\mathrm{curr}}(t) + \beta_3^s \zeta_{\mathrm{birth}}(\alpha,t)\zeta_{\mathrm{curr}}(t)}}$ |
| Environmental Saturation with Age-structure (ESA) | $G^s(\alpha,t) = \dfrac{e^{\beta_0^s \mathrm{logit}(s_0(\alpha)) + \beta_1^s \zeta_{\mathrm{birth}}(\alpha,t) + \beta_2^s \zeta_{\mathrm{curr}}(t) + \beta_4^s \zeta_{\mathrm{curr}}^2(t) + \beta_5^s \zeta_{\mathrm{birth}}(\alpha,t)\zeta_{\mathrm{curr}}^2(t)}}{1 + e^{e^{\beta_0^s \mathrm{logit}(s_0(\alpha)) + \beta_1^s \zeta_{\mathrm{birth}}(\alpha,t) + \beta_2^s \zeta_{\mathrm{curr}}(t) + \beta_4^s \zeta_{\mathrm{curr}}^2(t) + \beta_5^s \zeta_{\mathrm{birth}}(\alpha,t)\zeta_{\mathrm{curr}}^2(t)}}}$ |

Only the survival functions are displayed, however these functions equally apply to fecundity by swapping age-specific baseline functions and superscript indices. The realized survival $s(\alpha,t)$ is the product of the cohort effect function $G^s(\alpha,t)$ and the modifier function $g^s(\alpha)$ as given in equation 7b. The inverse-logit transform, $f(x) = e^x/(1+e^x)$, ensures realized vital rates stay on [0,1]. The parameters $\{\beta_i^j\}$ for $j \in \{f,s\}$ and $i \in \{0,1,2,3,4,5\}$ control the relative contribution of age-specific baselines, birth effects, current effects and any explicit interaction terms.

Table 2: Periodicity Metrics

| Metric | Category | Applies to | Calculation | Interpretation |
|---|---|---|---|---|
| $\boldsymbol{\rho}$ | Breadth | Whole model | Proportion of all runs that were periodic | Breadth of parameter space giving rise to or overall tendency of the scenario toward periodicity |
| $\boldsymbol{\psi}$ | Power | Individual run | The largest value of the global wavelet power across all periods with statistically significant global wavelet power | Strength of periodicity |
| $\boldsymbol{\Psi}$ | Power | Whole model | Highest of the $\psi$ across all periodic runs | Max strength of periodicity possible from the model |
| $\boldsymbol{\omega}$ | Period | Individual run | Period corresponding to $\psi$ | Dominant period |
| $\boldsymbol{\Omega}$ | Period | Whole model | $\omega$ of the run that produced $\Psi$ | Dominant period of the most strongly periodic run |
| $\boldsymbol{\tau}$ | Longevity | Individual run | Longest continuous stretch of years the *local* wavelet power associated with $\omega$ remained statistically significant | Duration of strongest periodic component |
| **T** | Longevity | Whole model | The $\tau$ of the run that produced $\Psi$ | Duration of strongest periodic component in the most strongly periodic run of the model |

Metrics of individual run and whole model output. Lowercase Greek letters refer to characteristics of individual runs or the average characteristics across all periodic runs of the same model. Uppercase Greek letters refer to the characteristics of the individual run with the highest significant global wavelet power of all runs in the same model; these maximum metrics, though technically about a single run, are used as representatives of the kind of periodicity possible under the model so we treated them as whole-model metric.

# Figure Legends

**Figure 1:** Periodicity metrics and cohort effect functional forms from simulation Rounds 1 (A) and 2 (B). (A) Periodicity metrics for density-dependent models under stochastic environmental scenarios. Icons correspond to models—environmental scenario (x-axis), density dependence hypothesis (color), and cohort effect hypothesis (shape)–showing proportion of periodic runs (a1), highest maximum power (a2), dominant period (a3) and longevity (a4). Red dashes (a1, a2) indicate Round 2 inclusion thresholds. (B) Round 2 cohort effect functional forms of parameters yielding $\psi$ in the top 10$^{th}$ percentile, showing median (line) and range (shaded) for different birth conditions (colors). Baselines (grey) are shown for comparison.

**Alt text:** Multipaneled figure. Left: four panels of scatterplots, each showing whole-model results for a different periodicity metric. Right: two panels of line graphs, each showing realized vital rates across different conditions of birth and adulthood; panels correspond to fecundity and survival respectively.

**Figure 2:** Sample run outputs from the June 20th GDD delayed density dependence SSA model. (A) Abundance (left) and average adult vital rates (right). (B) Local wavelet power spectrum; 90% significance threshold (white contour), power ridge (black), and longevity of the dominant period (τ, grey brackets) are shown. The first 100 years (greyed-out region) and COI (cone of influence, pink-shaded region) were excluded. (C) Cohort effect functional forms of parameters that produced the population trajectory in (A). Baselines (grey) are shown for comparison to realized rates by birth conditions (solid lines). (D) Fecundity (left) and survival (right) vs. 16-year lagged density.

**Alt text:** Multipaneled figure. Top left is a timeseries line graph of abundances and vital rates. Top right is an image plot of local wavelet power. Bottom left is a two-paneled line graph of realized vital rates for various conditions of birth and adulthood. Bottom right is a two-paneled scatterplot of realized vital rates by lagged density.

**Figure 3:** Sensitivity of periodicity metrics $\psi$ (strength, a1/b1) and $\tau$ (longevity, a2/b2) to vital rate coefficients under different environmental scenarios (A) and demographic assumptions (B). (A) Full density dependent SSA models across a sample of environmental scenarios. (B) Whole-year AO scenario across varied demographic assumptions. Bar height is the Sobol' index for first-order (solid) and total-order (striped) indices. Error bars (95% CI) are colored by statistical significance. Parameters $\{\beta_i^j\}, j \in \{f, s\}, i \in \{0,1,2\}$ denote age-specific baseline (0), birth effect (1) and current effect (2) coefficients for fecundity (f) and survival (s). Bar colors indicate the environmental scenario (A) or demographic scenario (B).

**Alt text:** Multipaneled figure. Each of the four panels is a barplot with 95% confidence intervals overlaid. Each panel shows first- and total-order Sobol' indices for each parameter for a given periodicity metric and model scenario.

**Figure 4:** High-level conceptual summary of the similarities and differences between long- and short-period population cycles. While diverse taxa can cycle due to similar kinds of feedbacks and interaction between time delays and demography, the nature of the relationships among drivers, the detectability, and the demographic characteristics fundamental to the occurrence of periodicity differ between rapid vs protracted cycles.
**Alt text:** Venn diagram comparison of long-period vs. short-period cycling mechanisms and characteristics.

Figure 1

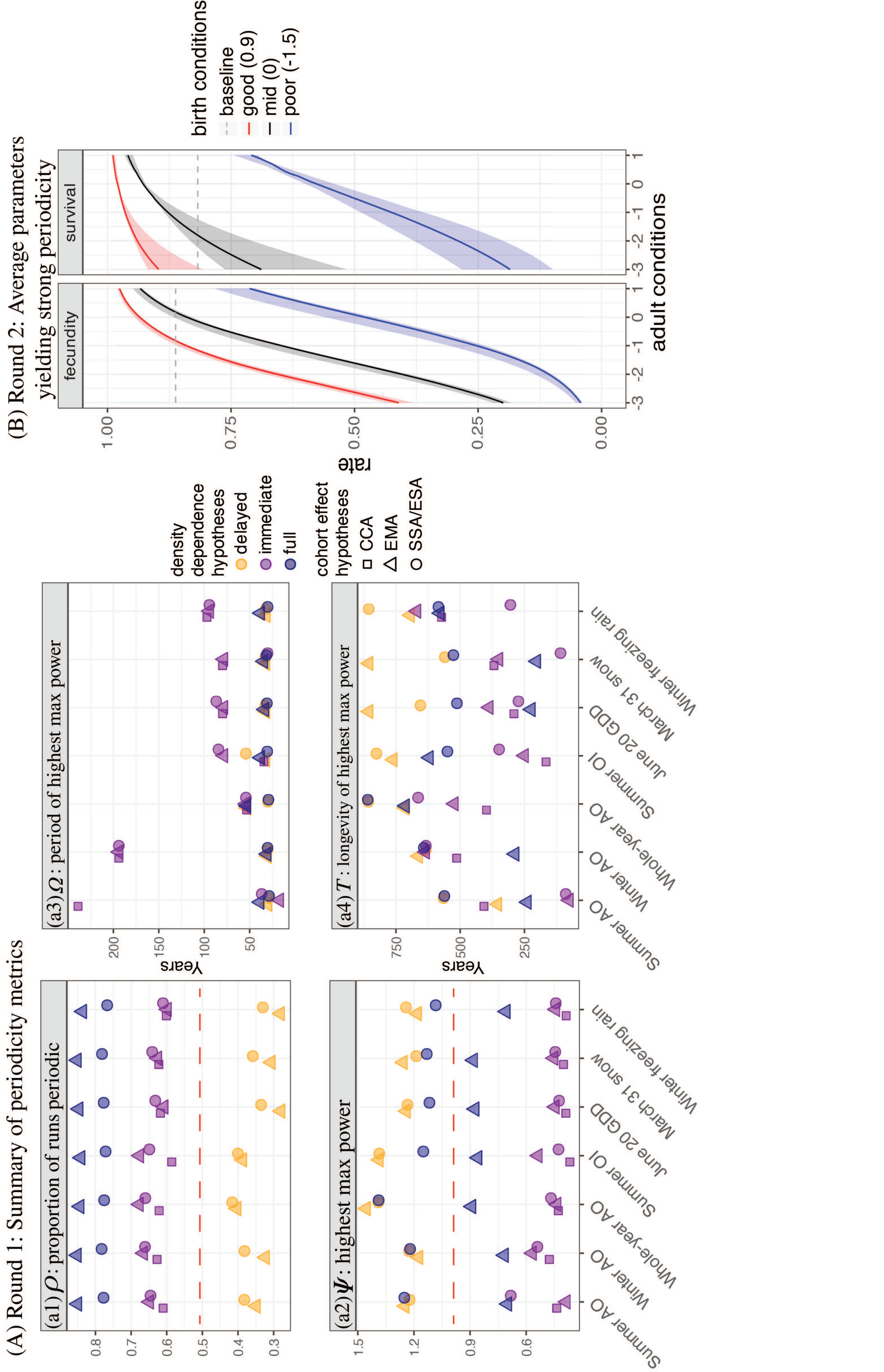
(A) Round 1: Summary of periodicity metrics
(a1) ρ : proportion of runs periodic
(a2) Ψ : highest max power
(a3) Ω : period of highest max power
(a4) T : longevity of highest max power
Years
Summer AO
Winter AO
Whole-year AO
Summer OI
June 20 GDD
March 31 snow
Winter freezing rain
density dependence hypotheses
delayed
immediate
full
cohort effect hypotheses
CCA
EMA
SSA/ESA
(B) Round 2: Average parameters yielding strong periodicity
fecundity
survival
rate
adult conditions
birth conditions
baseline
good (0.9)
mid (0)
poor (-1.5)

Figure 2

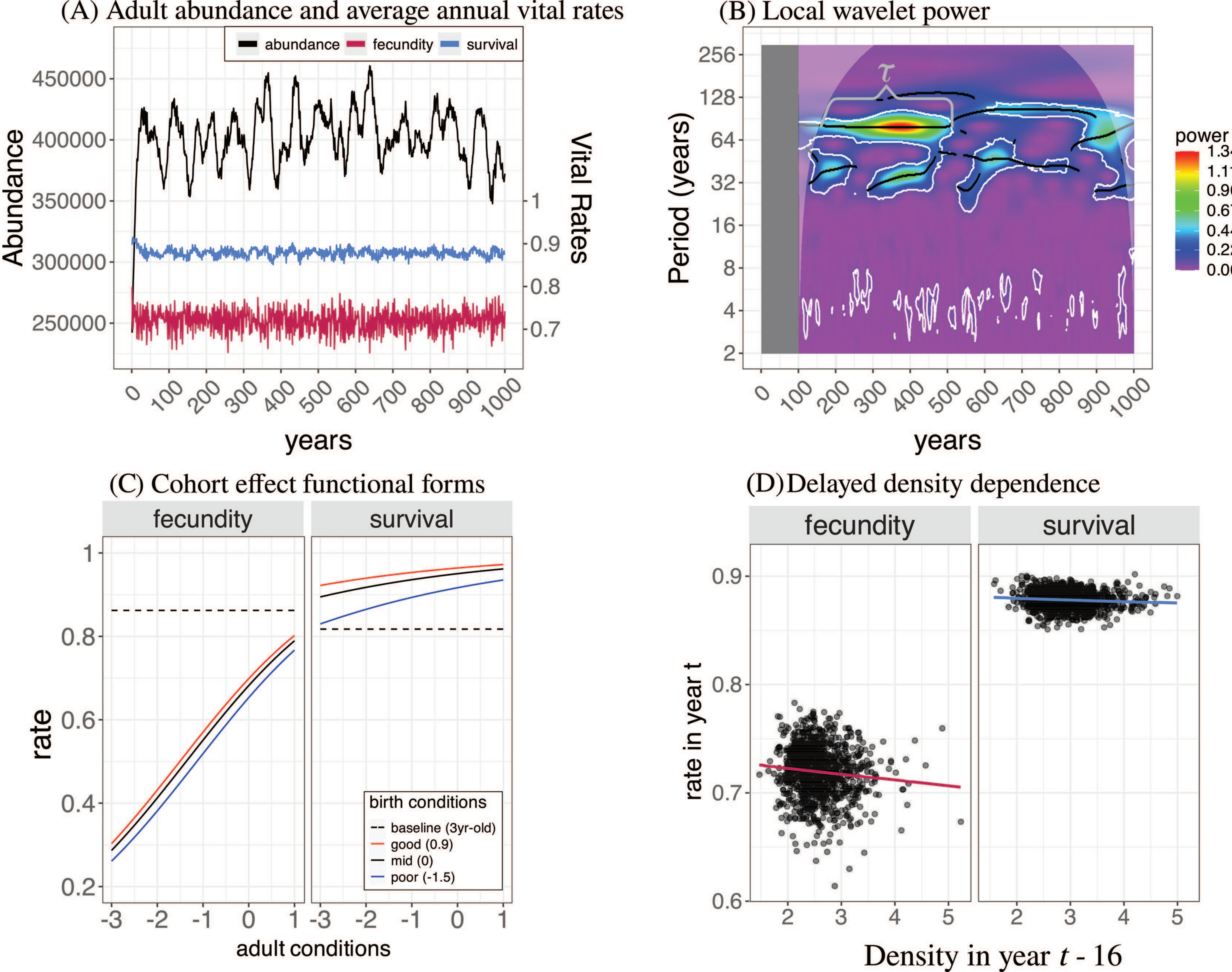
(A) Adult abundance and average annual vital rates
abundance
fecundity
survival
Abundance
450000
400000
350000
300000
250000
Vital Rates
1
0.9
0.8
0.7
0
100
200
300
400
500
600
700
800
900
1000
years
(B) Local wavelet power
τ
Period (years)
256
128
64
32
16
8
4
2
power
1.349
1.119
0.904
0.674
0.445
0.229
0.000
years
(C) Cohort effect functional forms
fecundity
survival
rate
1
0.8
0.6
0.4
0.2
-3
-2
-1
0
1
adult conditions
birth conditions
baseline (3yr-old)
good (0.9)
mid (0)
poor (-1.5)
(D) Delayed density dependence
fecundity
survival
rate in year t
0.9
0.8
0.7
0.6
2
3
4
5
Density in year t - 16

Figure 3

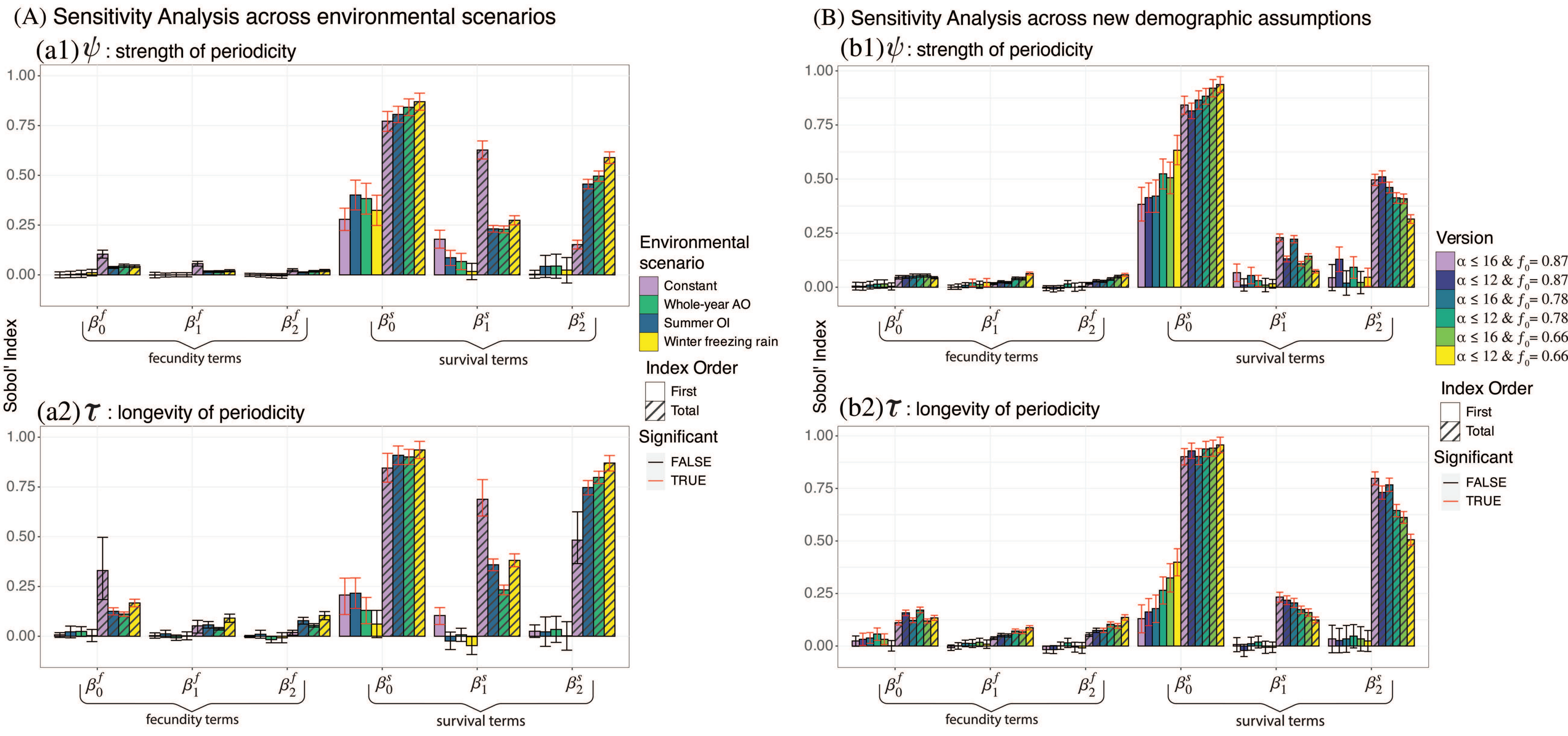

(A) Sensitivity Analysis across environmental scenarios
(a1) $\psi$ : strength of periodicity
(a2) $\tau$ : longevity of periodicity
Sobol' Index
1.00
0.75
0.50
0.25
0.00
$\beta_0^f$ $\beta_1^f$ $\beta_2^f$
fecundity terms
$\beta_0^s$ $\beta_1^s$ $\beta_2^s$
survival terms
Environmental scenario
Constant
Whole-year AO
Summer OI
Winter freezing rain
Index Order
First
Total
Significant
FALSE
TRUE
(B) Sensitivity Analysis across new demographic assumptions
(b1) $\psi$ : strength of periodicity
(b2) $\tau$ : longevity of periodicity
Version
$\alpha \leq 16$ & $f_0 = 0.87$
$\alpha \leq 12$ & $f_0 = 0.87$
$\alpha \leq 16$ & $f_0 = 0.78$
$\alpha \leq 12$ & $f_0 = 0.78$
$\alpha \leq 16$ & $f_0 = 0.66$
$\alpha \leq 12$ & $f_0 = 0.66$

Figure 4

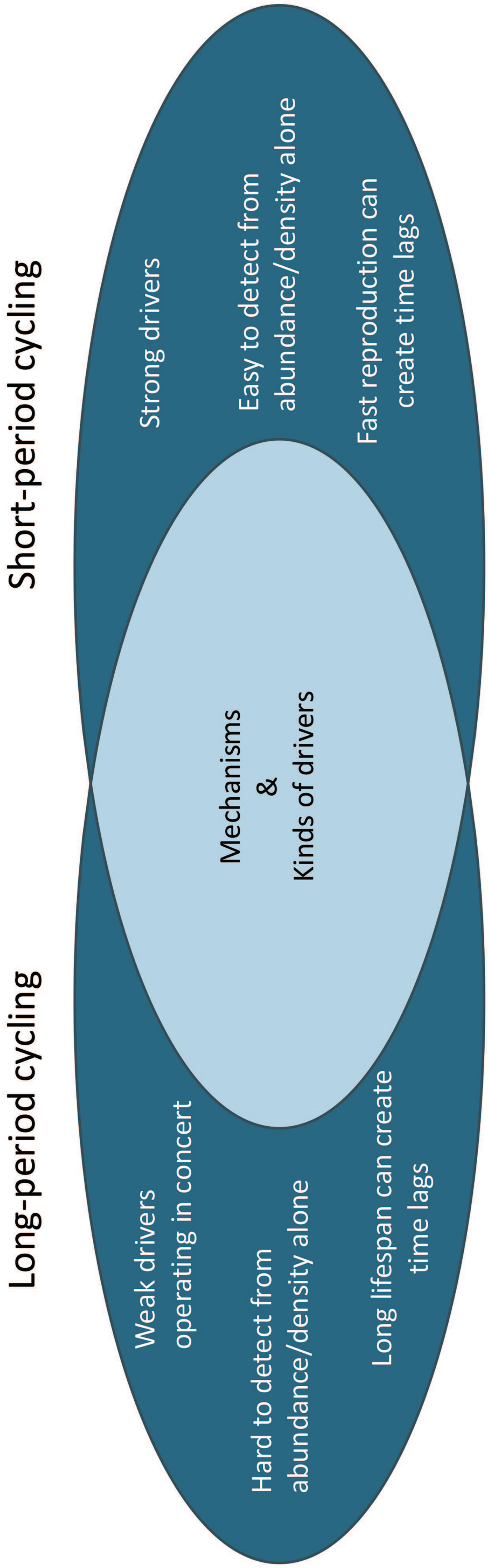

Short-period cycling
Strong drivers
Easy to detect from abundance/density alone
Fast reproduction can create time lags
Mechanisms
&
Kinds of drivers
Long-period cycling
Weak drivers operating in concert
Hard to detect from abundance/density alone
Long lifespan can create time lags

# Supplementary Material for: Drivers of periodicity in population dynamic models of long-lived, large mammals, *The American Naturalist*

Marron Clifford McConnell[1,*], William F. Fagan[1]

[1]Department of Biology, University of Maryland, College Park, MD, USA

[*]Corresponding author, email: mecmcc@umd.edu

# Table of Contents

# List of Figures

# List of Tables

# Appendix A: Functional forms of cohort effect hypotheses and exclusion of ESA forms from density-dependent models

All functional forms $G_h(\alpha, t)$ for cohort effect hypotheses are the inverse-logit transform (also known as the standard logistic function) of $l_h$, the sum of linear or quadratic terms with each expression specific to each hypothesis $h$: $G_h(\alpha, t) = \frac{e^{l_h(\alpha,t)}}{1+e^{l_h(\alpha,t)}}$. The inverse-logit transform ensures realized fecundity and survival cannot exceed 1.

Regardless of the hypothesis, the age-specific baseline term $f_0(\alpha)$ is first logit-transformed; this ensures the recovery of the age-specific baseline in the absence of other terms (i.e. in the absence of influence from current and birth conditions, the inverse-logit transform ought to return $f_0(\alpha)$). Each term of $l_h(\alpha, t)$ has a corresponding coefficient $\beta_i^j$ controlling the relative importance of each term in influencing the $j^{th}$ vital rate.

The Current Conditions with Age-structure hypothesis has the biological interpretation that realized fecundity and survival are impacted by age and the conditions of adulthood with no impact from the conditions of birth; as such, $l_{\text{CCA}}(\alpha, t)$ only has terms for the age-specific baseline and current conditions.

The Silver Spoon with Age-structure hypothesis has the biological interpretation that realized fecundity and survival are impacted by age, the conditions of adulthood, and the conditions of birth such that cohorts born during 'good' conditions always enjoy higher vital rates; accordingly, $l_{\text{SSA}}(\alpha, t)$ has terms for the age-specific baseline, current conditions, and birth conditions.

The Environmental Matching with Age-structure hypothesis has the biological interpretation that realized fecundity and survival are impacted by age, the conditions of adulthood, and the conditions of birth such that cohorts have the highest vital rates under adult conditions that match the conditions of their birth; therefore, in addition to the linear terms for age-specific baseline, current conditions and birth conditions, $l_{\text{EMA}}(\alpha, t)$ contains an additional term for the *interaction* between birth and current conditions. If the signs of $\zeta_{\text{birth}}$ and $\zeta_{\text{curr}}$ match, this interaction term is positive, thus increasing the vital rate. On the other hand, if the signs of $\zeta_{\text{birth}}$ and $\zeta_{\text{curr}}$ do not match, this interaction term is negative, thus decreasing the vital rate.

The Environmental Saturation with Age-structure hypothesis has the biological interpretation that realized fecundity and survival are impacted by age, conditions of adulthood and conditions of birth such that cohorts born in 'good' conditions enjoy higher vital rates only under intermediate adult conditions, and that the stratification by birth condition becomes negligible at both positive and negative extremes of current conditions. The influence of birth conditions should peak at an intermediate value of the adult conditions; if we took two cohorts $\alpha_m$ and $\alpha_n$ and required that $f_0(\alpha_m) = f_0(\alpha_n)$ and $\zeta_{\text{birth}}(\alpha_m, t) > \zeta_{\text{birth}}(\alpha_n, t)$, then plotting $l_{\text{ESA}}(\alpha_m, t) - l_{\text{ESA}}(\alpha_n, t)$ as a function of $\zeta_{\text{curr}}(t)$ should result in a concave down parabola. A negative effect of the interaction between birth condition and the square of current conditions ensures that the discrepancy among cohorts is constrained at large absolute values of current conditions; a negative effect of the square of current conditions determines at which values of $\zeta_{\text{curr}}(t)$ the saturation takes effect. However, the biological interpretation of this ESA functional form is only valid for constrained $\zeta(\cdot)$ values.

| Cohort Effect Hypothesis | Functional form for $G^s(\alpha,t)$ |
|---|---|
| Current Conditions with Age-structure (CCA) | $G^s(\alpha,t)=\frac{e^{\beta_0^s \mathrm{logit}(s_0(\alpha))+\beta_2^s\zeta_{\mathrm{curr}}(t)}}{1+e^{\beta_0^s \mathrm{logit}(s_0(\alpha))+\beta_2^s\zeta_{\mathrm{curr}}(t)}}$ |
| Silver Spoon with Age-structure (SSA) | $G^s(\alpha,t)=\frac{e^{\beta_0^s \mathrm{logit}(s_0(\alpha))+\beta_1^s\zeta_{\mathrm{birth}}(\alpha,t)+\beta_2^s\zeta_{\mathrm{curr}}(t)}}{1+e^{\beta_0^s \mathrm{logit}(s_0(\alpha))+\beta_1^s\zeta_{\mathrm{birth}}(\alpha,t)+\beta_2^s\zeta_{\mathrm{curr}}(t)}}$ |
| Environmental Matching with Age-structure (EMA) | $G^s(\alpha,t)=\frac{e^{\beta_0^s \mathrm{logit}(s_0(\alpha))+\beta_1^s\zeta_{\mathrm{birth}}(\alpha,t)+\beta_2^s\zeta_{\mathrm{curr}}(t)+\beta_3^s\zeta_{\mathrm{birth}}(\alpha,t)\zeta_{\mathrm{curr}}(t)}}{1+e^{\beta_0^s \mathrm{logit}(s_0(\alpha))+\beta_1^s\zeta_{\mathrm{birth}}(\alpha,t)+\beta_2^s\zeta_{\mathrm{curr}}(t)+\beta_3^s\zeta_{\mathrm{birth}}(\alpha,t)\zeta_{\mathrm{curr}}(t)}}$ |
| Environmental Saturation with Age-structure (ESA) | $G^s(\alpha,t)=\frac{e^{\beta_0^s \mathrm{logit}(s_0(\alpha))+\beta_1^s\zeta_{\mathrm{birth}}(\alpha,t)+\beta_2^s\zeta_{\mathrm{curr}}(t)+\beta_4^s\zeta_{\mathrm{curr}}^2(t)+\beta_5^s\zeta_{\mathrm{birth}}(\alpha,t)\zeta_{\mathrm{curr}}^2(t)}}{1+e^{e^{\beta_0^s \mathrm{logit}(s_0(\alpha))+\beta_1^s\zeta_{\mathrm{birth}}(\alpha,t)+\beta_2^s\zeta_{\mathrm{curr}}(t)+\beta_4^s\zeta_{\mathrm{curr}}^2(t)+\beta_5^s\zeta_{\mathrm{birth}}(\alpha,t)\zeta_{\mathrm{curr}}^2(t)}}}$ |

Table A-1: Copy of Table 1 in the main text. Functional forms governing vital rates corresponding to each hypothesis of cohort effects. Only the survival functions are displayed, however these functions equally apply to fecundity by swapping age-specific baseline functions and superscript indices. The realized survival $s(\alpha,t)$ is the product of the cohort effect function $G^s(\alpha,t)$ and the modifier function $g^s(\alpha)$ as given in equation 7b. The inverse-logit transform, $f(x)=\frac{e^x}{1+e^x}$, ensures realized vital rates stay on [0,1]. The parameters $\{\beta_i^j\}$ for $j\in\{f,s\}$ and $i\in\{0,1,2,3,4,5\}$ control the relative contribution of age-specific baselines, birth effects, current effects and any explicit interaction terms.

For models without density dependence, birth effects $\zeta_{\mathrm{birth}}(\alpha,t)=\varepsilon(t-\alpha)$, and current effects $\zeta_{\mathrm{curr}}(t)=\varepsilon(t)$ are bounded within the closed interval $[-1,1]$ because of the rescaling of $\varepsilon(t)$ (environmental scenario time series). This provides a clear range of sensible parameter sets $\{\beta_i^j\}$ that align with the biological interpretation of the various hypotheses. However, for density dependent models, the $\zeta(\cdot)$ are functions of time, age, and density: $\zeta_{\mathrm{birth}}(\alpha,t)=1-\frac{A(t-\alpha)}{K(t-\alpha)}$ and $\zeta_{\mathrm{curr}}(t)=1-\frac{A(t)}{K(t)}$. For these models, the lower bound of $A(t)$ provides an upper bound for $\zeta(t)$ because $\forall\ t, A(t)\geq 0$ and thus $1-\frac{A(t)}{K(t)}\leq 1$. However, $A(t)$ is not bounded above and thus there is no lower bound for $1-\frac{A(t)}{K(t)}$. For $\zeta(\cdot)$ values much below -1, the environmental saturation with age structure (ESA) hypothesis has a vanishingly narrow range of parameter values for which the functional form still matches the biological interpretation. Fortunately, there is a redundancy in the ESA form due to the application of the inverse-logit function to the entire expression: the saturating nature of the inverse-logit function already reduces the discrepancies between extreme values while better preserving discrepancies for intermediate values. One can therefore recover an ESA-like functional form with saturation at extremes from the SSA equation when $\beta_0$ (coefficient of age-baseline term) and $\beta_2$ (coefficient of current effects term) are sufficiently large. For these reasons, the ESA form is only explicitly considered for non-density-dependent models. For density-dependent models, the hypothesis of environmental saturation is subsumed in the SSA functional form, and the biological distinction between silver spoon and saturation effects can be assessed by comparing the difference between "good" and "poor" birth outcomes under extreme conditions with the same difference under intermediate conditions. If the difference increases under intermediate conditions, this corresponds to the ESA hypothesis, whereas if the difference decreases or does not change under intermediate conditions, this corresponds to the SSA hypothesis.

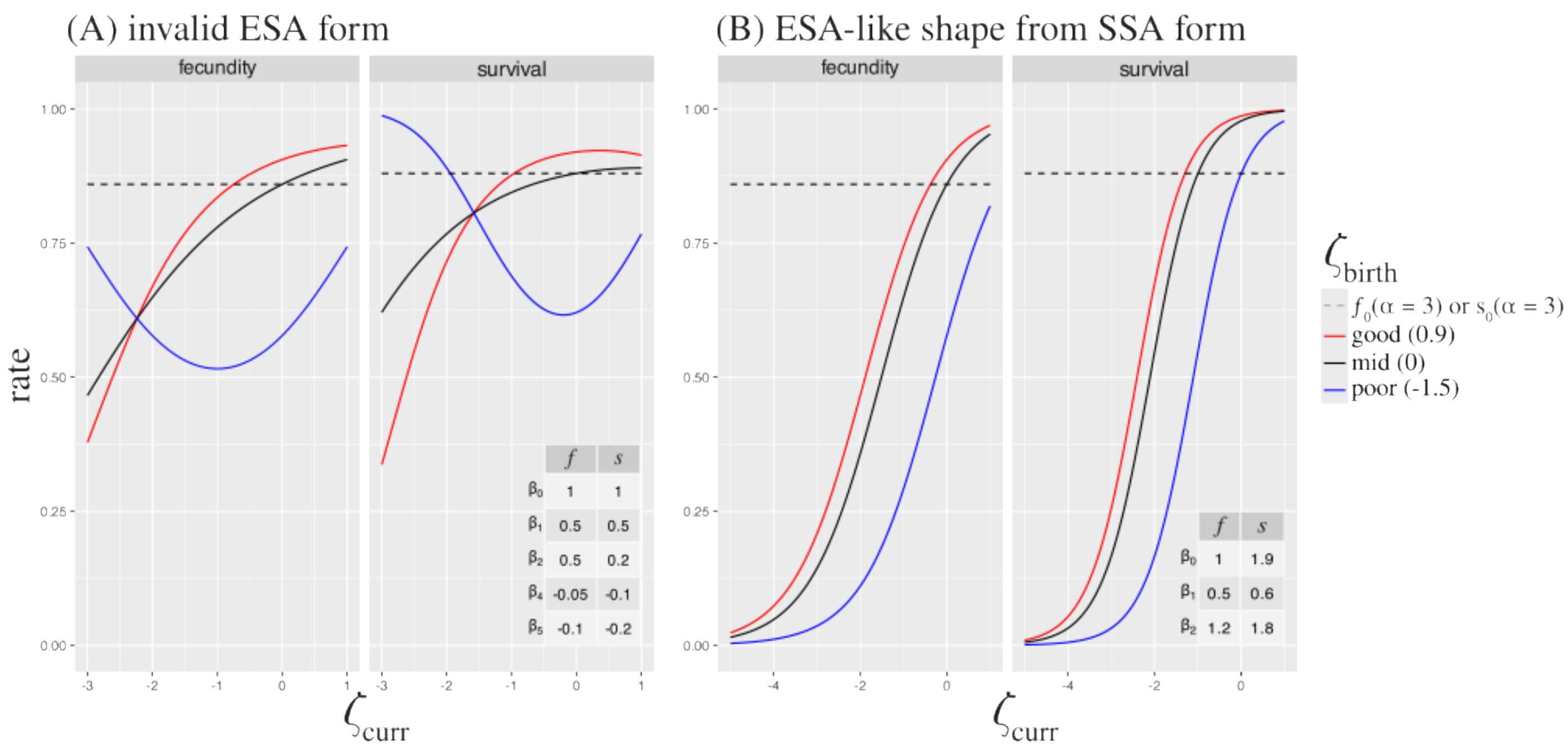

Figure A-1: Example of ESA form (A) and SSA form (B) with reasonable coefficient values, illustrating cohort effect hypothesis curves when $\zeta_{\text{birth}}(\alpha, t)$ and $\zeta_{\text{curr}}(t)$ values are not bounded below by $-1$. The biological interpretation of saturation at extremes of adult conditions no longer aligns with the environmental saturation with age-structure (ESA) functional form for adult condition values below $-1$, but for some values, does align with the silver spoon with age-structure (SSA) functional form, despite its lack of explicit quadratic terms. For $\zeta(\cdot)$ with no lower bound, the SSA functional form can encapsulate both the silver spoon and environmental saturation hypotheses.

Across the four hypotheses of cohort effect and four hypotheses of density dependence, there are 16 total combinations. Of these, three ESA models were invalid per the above. There were additionally two redundant combinations of density dependence hypotheses for CCA models. In all, that left 11 unique and valid combinations of cohort effect and density dependence hypotheses, each with 8 environmental scenarios, for a total of 88 unique and valid models.

| | **Cohort effect hypotheses** | | | |
|---|---|---|---|---|
| **Density dependence hypotheses** | **Current Conditions with Age-structure (CCA)** | **Silver Spoon with Age-structure (SSA)** | **Environmental Matching with Age-structure (EMA)** | **Environmental Saturation with Age-structure (ESA)** |
| **Environment Only** | Constant scenario + 7 empirically motivated scenarios | Constant scenario + 7 empirically motivated scenarios | Constant scenario + 7 empirically motivated scenarios | Constant scenario + 7 empirically motivated scenarios |
| **Delayed density dependence** | NA (not unique, same as EO CCA model) | Constant scenario + 7 empirically motivated scenarios | Constant scenario + 7 empirically motivated scenarios | NA (not valid, see SI Appendix A) |
| **Immediate density dependence** | Constant scenario + 7 empirically motivated scenarios | Constant scenario + 7 empirically motivated scenarios | Constant scenario + 7 empirically motivated scenarios | NA (not valid, see SI Appendix A) |
| **Full density dependence** | NA (not unique, same as immediate density dependence CCA) | Constant scenario + 7 empirically motivated scenarios | Constant scenario + 7 empirically motivated scenarios | NA (not valid, see SI Appendix A) |

Table A-2: All possible combinations of cohort effect hypotheses (columns), density dependence (rows), and environmental scenarios (entries). There are a total of 88 unique and valid triplets necessary to constitute a model.

# Appendix B: Environmental scenarios and robustness of results

## B.1 Details of environmental scenarios

The Arctic Oscillation (AO) and Pacific Decadal Oscillation (PDO) are large-scale ocean-driven climate indices that act with varying intensity across longitudinal gradients in the Arctic and northern boreal zones of North America, as drivers of weather patterns including temperature and precipitation (National Oceanic and Atmospheric Administration National Centers for Environmental Information Climate Monitoring 2023). Across multiple herds spanning the North American continent, these indices have been linked not only to aspects of range conditions (Post and Stenseth 1999; Putkonen and Roe 2003; Mallory et al. 2018), but also to herd behavior (Gurarie et al. 2019; Couriot et al. 2023) and fitness and population dynamics (Aanes et al. 2002; Forchhammer et al. 2002; Griffith et al. 2002; Couturier et al. 2009; Joly et al. 2011; English et al. 2017; Hansen et al. 2019). We elected to explore the AO, for which the dynamics of herds in central Canada show closer association (Mallory et al. 2018). Herd range conditions, such as temperature, precipitation and vegetation, have also been linked to fluctuations in both survival and fecundity (Boulanger and Adamczewski 2017; Adamczewski et al. 2021; Boulanger et al. 2021). Boulanger and Adamczewski (2017) identified several of these metrics as more closely linked to demography, and we selected seven such metrics as our final set of environmental drivers: the Summer, Winter, and Whole-year averages of AO indices, the Summer Oestrid Index (OI, a composite metric of temperature and wind as a proxy for the severity of insect harassment during the summer, Russell et al. 2013), June 20$^{th}$ growing-degree-days (GDD), March 31$^{st}$ snow depth and Winter freezing rain.

| Scenario | Interpretation | Years | Season | Fecundity Correlation | Survival Correlation | Source: |
|---|---|---|---|---|---|---|
| Winter AO | Arctic Oscillation index avg: Jan-March | 1950-2024 | Winter | + | + | NOAA |
| Summer AO | Arctic Oscillation index avg: June-Aug | 1950-2024 | Summer | + | + | NOAA |
| Whole-year AO | Arctic Oscillation index avg: whole year | 1950-2024 | Year | + | + | NOAA |
| Summer OI | Cumulative Oestrid Index: June-Aug | 1980-2022 | Summer | - | - | MERRA |
| June 20 GDD | Cumulative growing degree days over 5°C by June 20th | 1980-2022 | Late spring/early summer | - | + | MERRA |
| March 31 snow | Snow depth on March 31st | 1980-2022 | Late winter/onset of spring migration | - | + | MERRA |
| Winter freezing rain | Cumulative freezing rain | 1980-2022 | Winter | - | - | MERRA |

Table B-1: Simulated environmental scenarios. Correlations between environmental values and demographic rates were termed positive (+) or negative (-) based on their net effects on demographic rates according to integrated population modelling of the Bathurst herd (Boulanger and Adamczewski 2017). Auto-regressive integrated moving average (ARIMA) models were fit to these time series, and new time series simulated from model fits. Data were sourced from the National Oceanic and Atmospheric Administration (NOAA) and the Modern Era Retrospective Analysis for Research and Applications (MERRA) database maintained by the CircumArctic Rangifer Monitoring Association (CARMA, Russell et al. 2013).

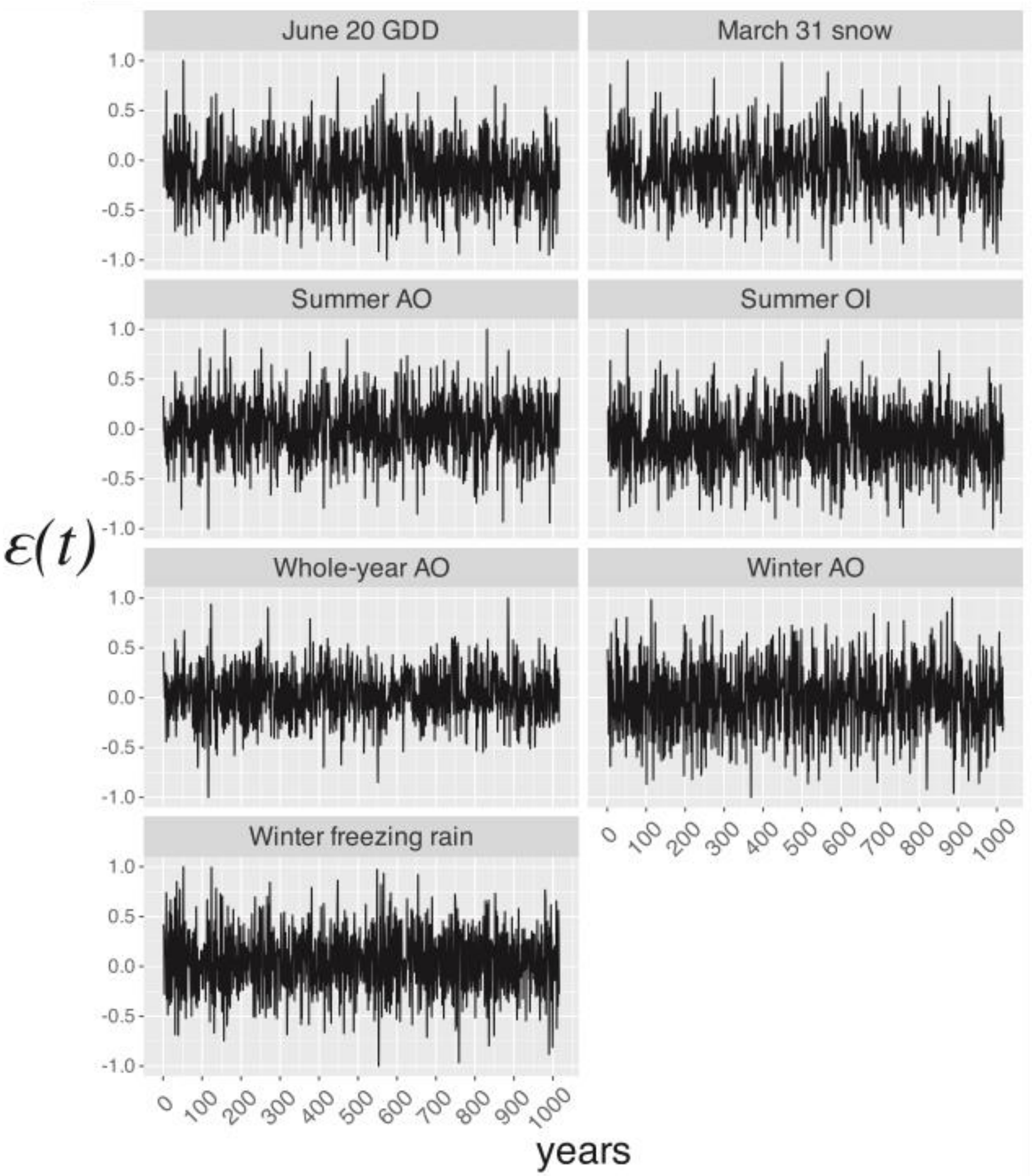

Figure B-1: Timeseries of empirically motivated environmental scenarios used for the first two rounds of simulation. Timeseries $\varepsilon(t)$ were simulated from ARIMA model fits to empirical time series and subsequently rescaled such that $\varepsilon(t) \in [-1,1]$.

## B.2 Robustness of qualitative results to specific realizations of environmental scenario

We fixed a set of parameters, $p_{avg}$, for each model as follows. Out of the set of 30,000 runs for each model from round 2, we took the distribution of all $\psi$ (strength of periodicity per run), selected only those runs with $\psi$ at or above the 90th percentile, and then calculated the average of each parameter. We also fixed a second set of parameters, $p_{top}$, for each model as follows. Out of all 30,000 runs for each model from round 2, we selected the single run that produced $\Psi$ (maximum strength of periodicity per model) and fixed the parameters $\{\beta_i^j\}$ that produced that run. Across the seven models, we found that these $p_{top}$ sets represented only four parameter sets: the first set was responsible for producing $\Psi$ in models with the four MERRA-based scenarios (Summer OI, June 20th GDD, Winter freezing rain, and March 31st snow) while the remaining three sets each corresponded to models with one of the AO-based scenarios.

We generated an additional 5,000 environmental timeseries $\varepsilon(t)$ for each scenario to confirm that periodicity arising from the previous rounds for these parameter sets were not simply flukes of the particular realizations of each environmental scenario. With these new $\varepsilon(t)$, we simulated 10,000 population trajectories for each of the seven models (5000 runs using the $p_{top}$ and 5000 runs using the $p_{avg}$ parameter sets for each scenario). All 70,000 runs – across all scenarios and both parameter sets—exhibited substantial periodicity. Given this, it appears likely that such parameter sets can drive periodicity across numerous environmental scenarios.

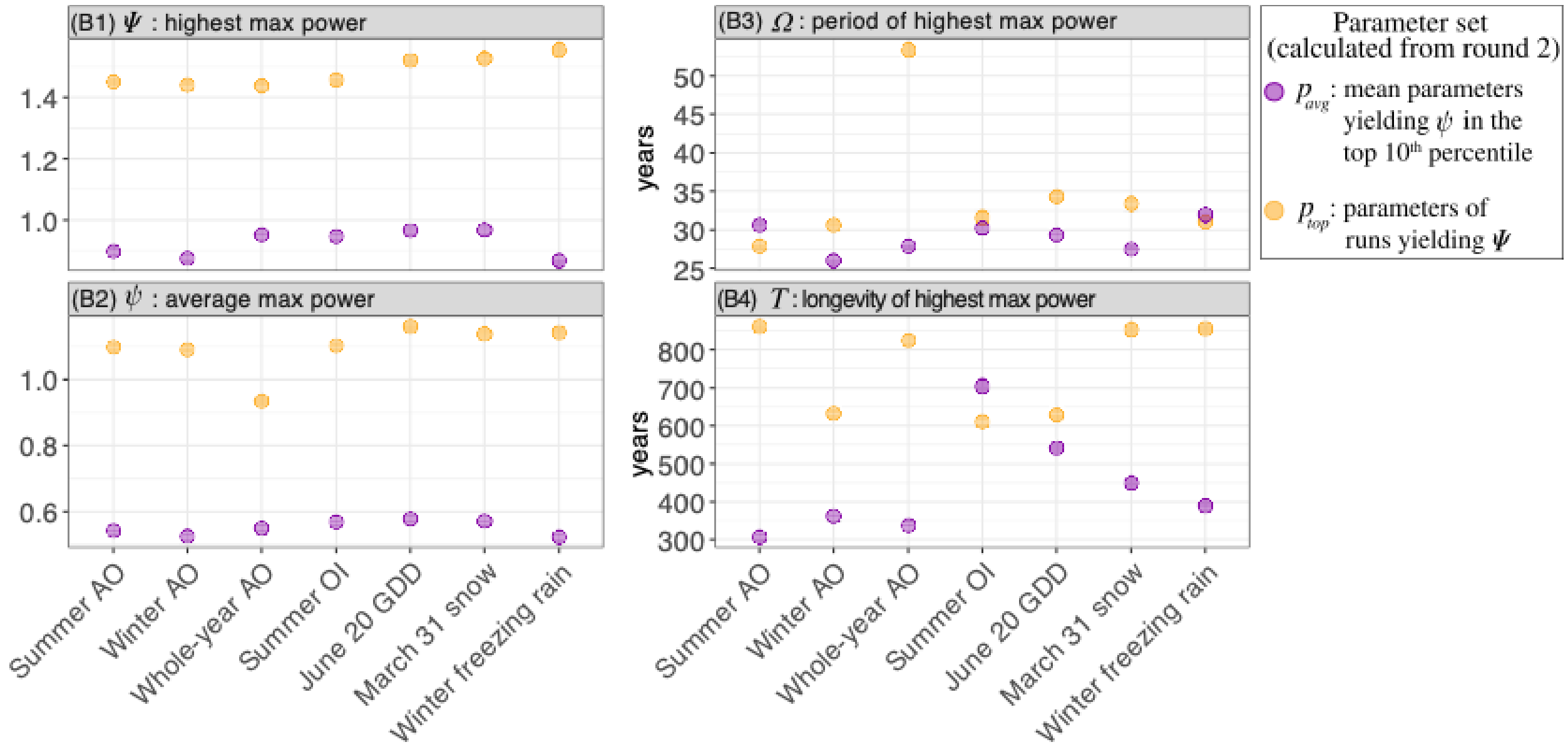

Figure B-2: Whole model summary metrics from simulation round 3 for the reduced set of models with fixed parameter sets. (B1) While models run with $p_{top}$ parameters outperformed their $p_{avg}$ counterparts, models run with $p_{avg}$ parameters still yielded strong periodicity. (B2) Although $\bar{\psi}$ values were roughly twice as high with $p_{top}$ as with $p_{avg}$ parameters, the $p_{avg}$ runs still produced remarkably high values. (B3) Dominant periods were almost always between 25-35 years. (B4) The most strongly periodic run of every model had continuous significant oscillations for a minimum of 300 years.

# Appendix C: Detailed results from simulation rounds 1 and 2

## C.1 Wavelet-analysis-based metrics

| Metric | Category | Applies to | Calculation | Interpretation |
|---|---|---|---|---|
| $\rho$ | Breadth | Whole model | Proportion of all runs that were periodic | Breadth of parameter space giving rise to or overall tendency of the scenario toward periodicity |
| $\psi$ | Power | Individual run | the largest value of the global wavelet power across all periods with statistically significant global wavelet power | Strength of periodicity |
| $\bar{\psi}$ | Power | Whole model | Average $\psi$ across all periodic runs | Strength of periodicity of an average run of the model |
| $\Psi$ | Power | Whole model | Highest of the $\psi$ across all periodic runs | Max strength of periodicity possible from the model |
| $\omega$ | Period | Individual run | Period corresponding to $\psi$ | Dominant period |
| $\bar{\omega}$ | Period | Whole model | Average $\omega$ across all periodic runs | Average dominant period |
| $\Omega$ | Period | Whole model | $\omega$ of the run that produced $\Psi$ | Dominant period of the most strongly periodic run |
| $\tau$ | Longevity | Individual run | Longest continuous stretch of years that the local wavelet power associated with $\omega$ remained statistically significant | Lifespan of strongest periodic component |
| $\bar{\tau}$ | Longevity | Whole model | The average of $\tau$ across all periodic runs | Average lifespan of strongest periodic components |
| T | Longevity | Whole model | The $\tau$ of the run that produced $\Psi$ | Lifespan of strongest periodic component in the most strongly periodic run of the model |

Table C-1:Copy of Table 3 in the main text explaining metrics of individual run and whole model output from wavelet analysis. Lowercase Greek letters refer to characteristics of individual runs or the average characteristics across all periodic runs of the same model. Uppercase Greek letters refer to the characteristics of the individual run with the highest significant global wavelet power of all runs in the same model; these maximum metrics, though technically about a single run, are used as representatives of the kind of periodicity possible under the model so we treated them as whole-model metrics. Expressions with an overscript bar refer to average characteristics of periodic runs of the same model.

## C.2 Constant scenario results

We analyzed all four cohort effect hypotheses and all four density dependence hypotheses under an environmental scenario of constant zeros to explore model performance in the absence of stochastic environmental drivers to better contextualize model performance under the seven empirically motivated scenarios. For the three classes incorporating density dependence, this is the equivalent to having a fixed carrying capacity independent of interannual environmental variation. Constant environmental scenario models were explored in the first (model elimination) and second (full density dependence SSA sensitivity analysis) rounds of simulation.

In the first round, none of the environment-only, immediate density dependence, nor CCA models resulted in periodicity for even a single run, and the full density dependence EMA model had less than 2% of runs result in periodicity ($\rho < 0.02$). The remaining models were comparable in terms of all metrics, however only the full density dependence SSA model was carried into the second round, as it exhibited the strongest wavelet power (the highest $\Psi$ at 1.67).

In the second round, the population trajectory created from the $p_{top}$ set of parameters exhibited sustained oscillations for the entire thousand-year span of simulation with a dominant period of 29 years. However, very few periodic runs exhibited sustained oscillations lasting the entire thousand years: less than 1% of runs exhibited periodicity lasting more than 100 years and only 10 runs exhibited undamped periodicity lasting the entire 1000 years. The majority of periodic runs exhibited damped oscillations approaching a fixed equilibrium.

Overall, the constant scenario was capable of producing strong oscillations in population trajectories but only when the model form was Silver Spoon effect with Age-structure (SSA) or Environmental Matching with Age-structure (EMA) and the model class was delayed or full density dependence; even then, periodicity emerged only with a limited number of parameter combinations The periodic runs under the constant scenario exhibited sustained oscillations lasting the full 1000 years of simulation, but for most, the oscillations were damped. That is, in the absence of additional ongoing perturbations, these cycles would fade away over time.

We also subjected the constant scenario full density dependence SSA model to Sobol' sensitivity analysis (Figure 1 in the main text). When the output metric was $\tau$ (longevity), no Sobol' indices, neither first- nor total-order, for fecundity coefficients were significant. First- and total-order indices for both $\beta_0^s$ and $\beta_1^s$ were significant, but no index for $\beta_2^s$ was significant. Fully 35% of the variance in $\tau$ was attributable to first-order effects. When the output metric was $\psi$ (strength), again no Sobol' indices for fecundity coefficients were significant, but all total-order indices for survival coefficients and first-order indices for $\beta_0^s$ and $\beta_1^s$ were significant. In this case, 46% of the variance in $\psi$ was attributable to first-order effects.

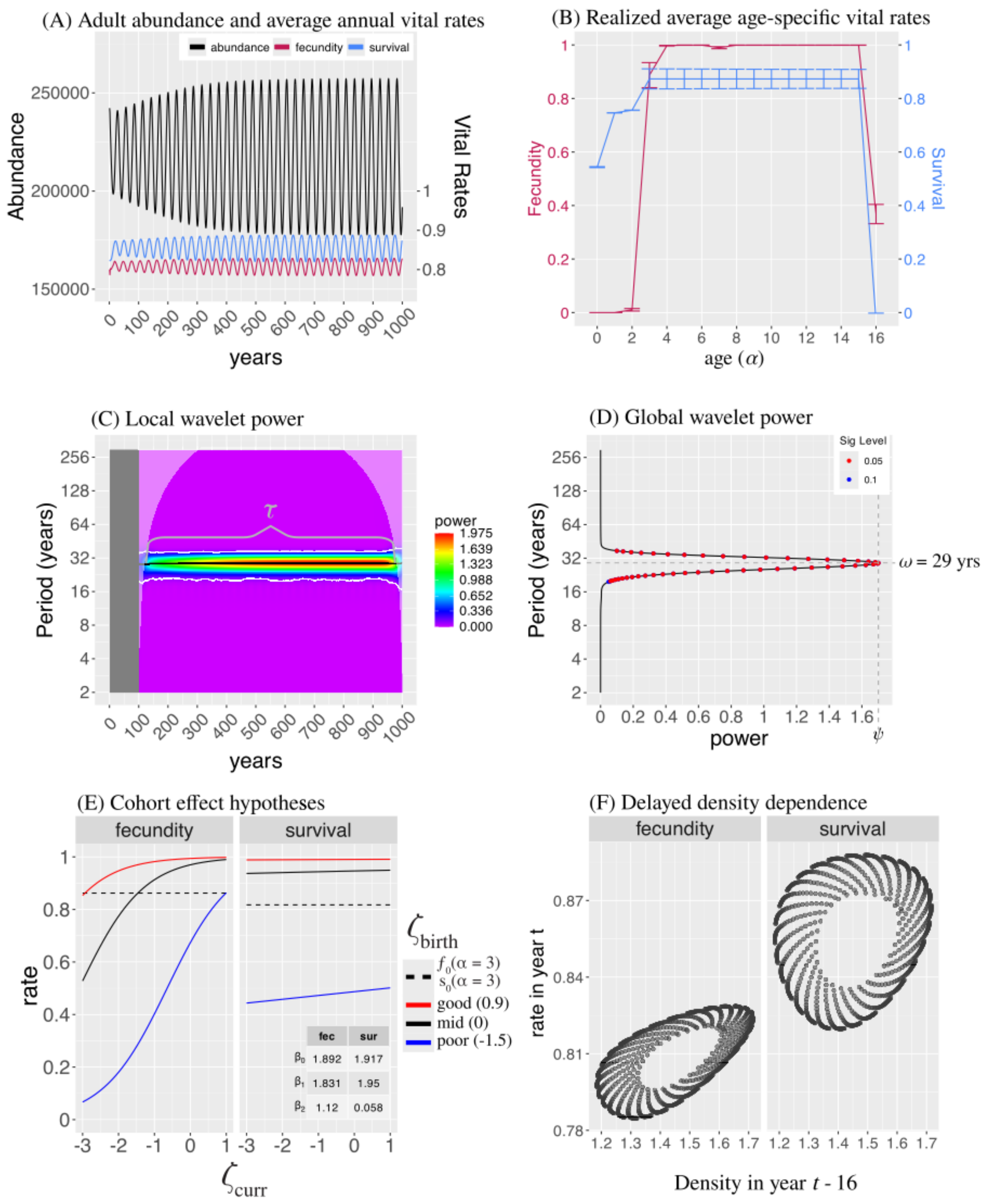

Figure C-1: Abundance, demographic rates, wavelet analysis outputs and hypotheses of cohort effects and density dependence from the Constant full density dependence SSA model run with $p_{top}$ parameters. (A) Adult female abundance (left) and average adult vital rates (right). (B) Average (±SD) age-specific fecundity (left y-axis) and survival (right y-axis) across all 1,000 years of simulation. (C) Local wavelet power spectrum with 90% significance threshold outlined (white) and ridge (black) indicating local peaks in power. The first 100 years (greyed-out portion) were trimmed prior to wavelet analysis. The COI (cone of influence, (Cazelles et al. 2008), light

pink-shaded boundary region) was excluded from interpretation. Longevity ($\tau$, grey brackets) of the dominant period is marked. (D) Global average wavelet power with $\psi$ (peak power) and $\omega$ (period) marked. (E) Cohort effect hypotheses for a three year-old based on parameter values $\{\beta_i^j\}$ that produced the run that yielded $\Psi$ (highest global wavelet power). Baselines are given by dashed lines and realized rates with three different birth conditions are given by solid colored lines. Parameter values that created these curves are listed in the table inset. (F) Relationship of fecundity (left, red trend line) and survival (right, blue trend line) to 16-year lagged density. Density in a given year was calculated as $A(t)/K(t)$, the abundance of adult females in a given year divided by the time-dependent carrying capacity in a given year.

## C.3 Round 1 results overview

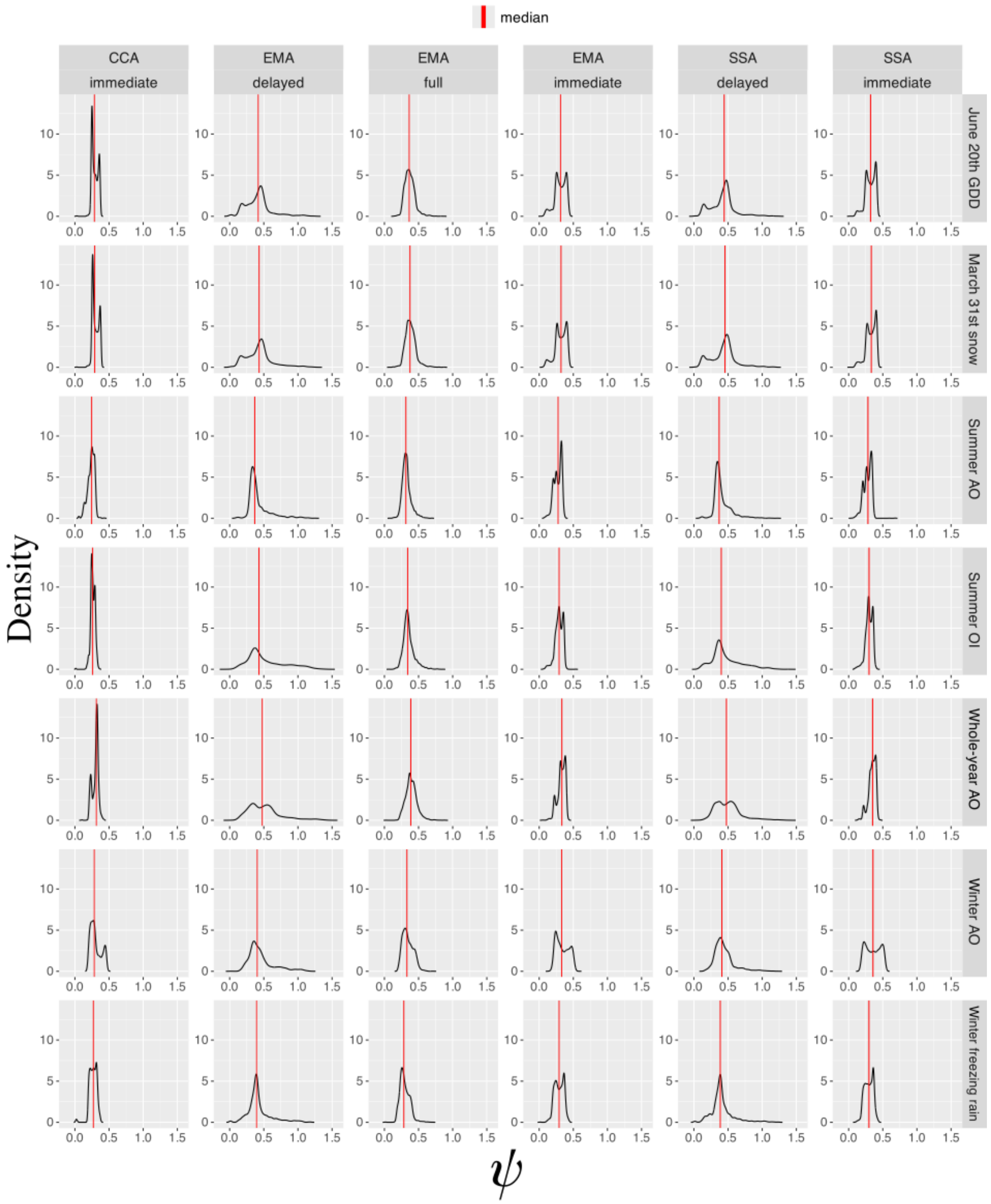

Figure C-2: Distributions of $\psi$ with medians (red) from only periodic runs of models eliminated after the first round of simulation. Models without density dependence had too few periodic runs to generate meaningful density curves. The distributions for EMA models strongly depended on model class: delayed model class distributions exhibited wider distributions with long rightward tails, immediate model class distributions exhibited jaggedness resembling CCA models, and full model class distributions fell in between. The distributions of SSA models closely resembled their same-class EMA counterparts. Across the seven scenarios, medians are highly conserved within each pair of model form and model class.

## C.4 Round 1 results detail: example of population trajectories

### C.4.1 Episodically strong or consistently intermediate periodicity

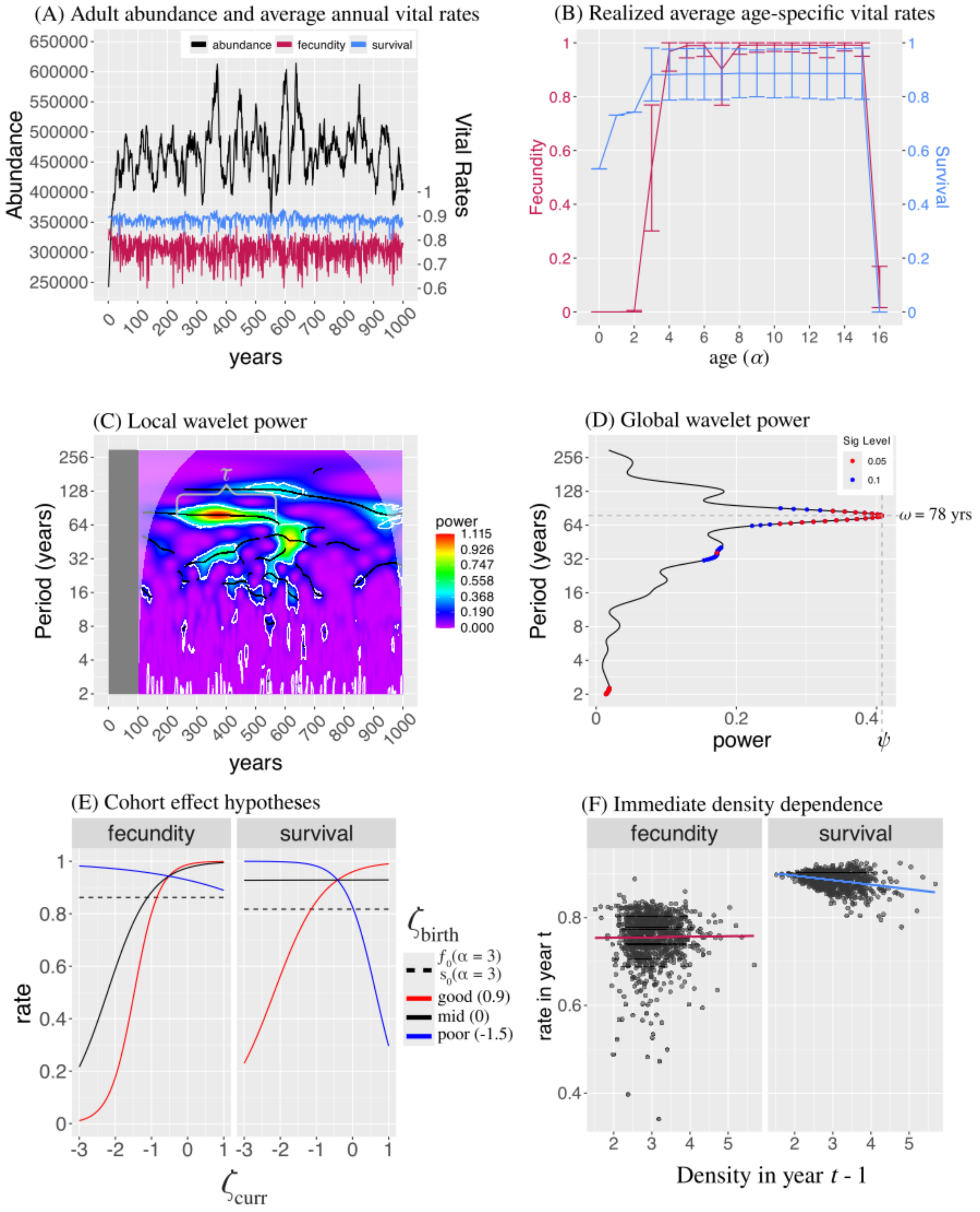

Figure C-3: Abundance, demographic rates, wavelet analysis outputs and hypotheses of cohort effects and density dependence from a sample run of the Summer OI immediate density dependence EMA model. (A) Adult female abundance (left) and average adult vital

rates (right). (B) Average (±SD) age-specific fecundity (left y-axis) and survival (right y-axis) across all 1,000 years of simulation. (C) Local wavelet power spectrum with 90% significance threshold outlined (white) and ridge (black) indicating local peaks in power. The first 100 years (greyed-out portion) were trimmed prior to wavelet analysis. The COI (cone of influence, (Cazelles et al. 2008), light pink-shaded boundary region) was excluded from interpretation. Longevity ($\tau$, grey brackets) of the dominant period is marked. (D) Global average wavelet power with $\psi$ (peak power) and $\omega$ (period) marked. Despite large changes in abundance, fluctuations were noisy, irregular and discontinuous enough to result in only moderate global wavelet power. (E) Cohort effect hypotheses for a three year-old based on parameter values $\{\beta_i^j\}$ that produced the population trajectory in (A). Baselines are given by dashed lines and realized rates with three different birth conditions are given by solid colored lines. Parameter values that created these curves are listed in the table inset. (F) Relationship of fecundity (left, red trend line) and survival (right, blue trend line) to density of the previous year. Density in a given year was calculated as ${A(t)}/{K(t)}$, the abundance of adult females in a given year divided by the time-dependent carrying capacity in a given year.

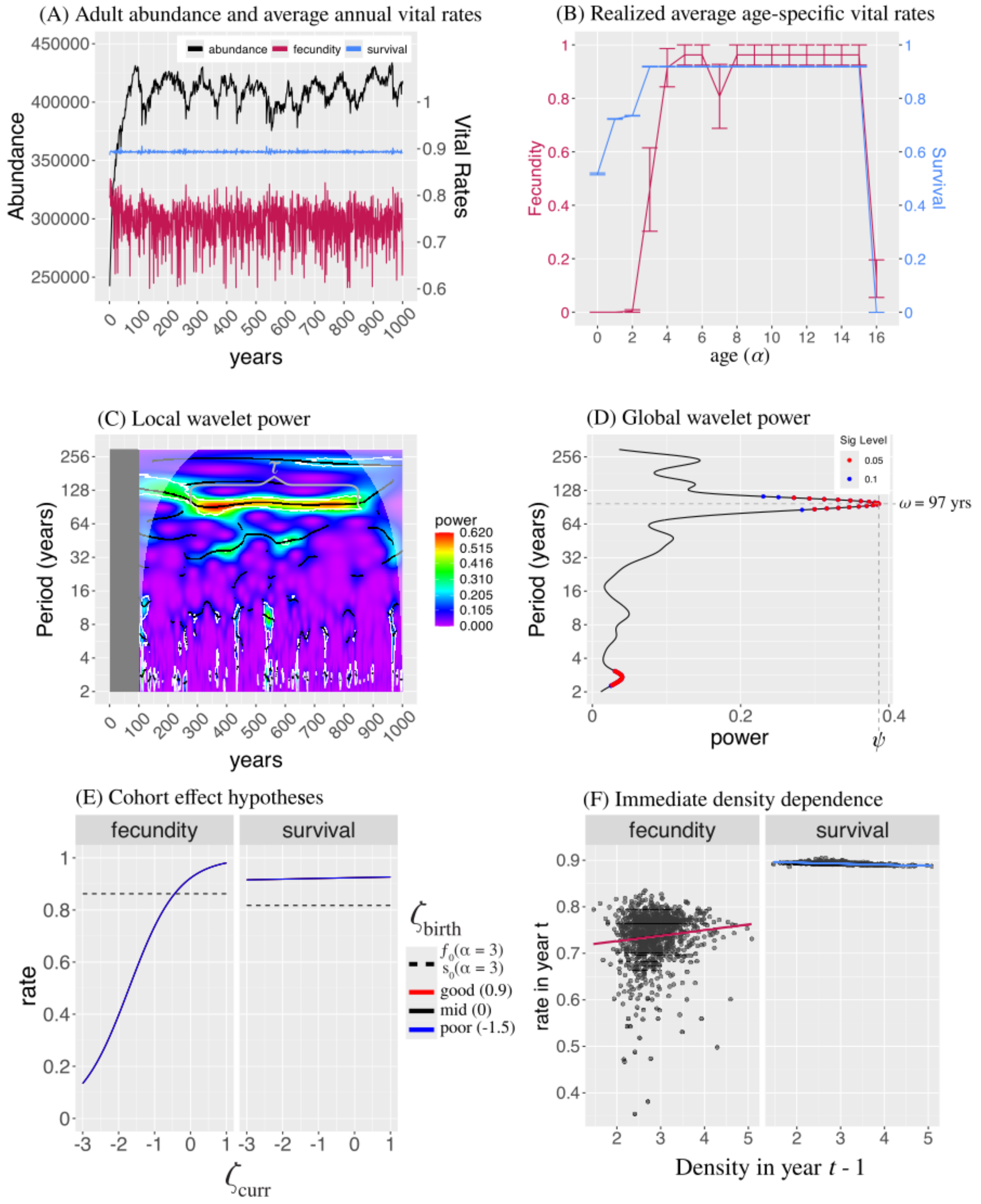

Figure C-4: Abundance, demographic rates, wavelet analysis outputs and hypotheses of cohort effects and density dependence from a sample run of the Winter freezing rain immediate density dependence CCA model. (A) Adult female abundance (left) and average adult vital rates (right). (B) Average (±SD) age-specific fecundity (left y-axis) and survival (right y-axis) across all 1,000 years of simulation. (C) Local wavelet power spectrum with 90% significance threshold outlined (white) and ridge (black) indicating local peaks in power. The first 100 years (greyed-out portion) were trimmed prior to wavelet analysis. The COI (cone of influence, (Cazelles et al.

2008), light pink-shaded boundary region) was excluded from interpretation. Longevity ($\tau$, grey brackets) of the dominant period is marked. (D) Global average wavelet power with $\psi$ (peak power) and $\omega$ (period) marked. The fluctuations in abundance are noisy and although the abundance timeseries is periodic for the entire duration of the simulation, it is never strongly periodic and so only results in moderately low global wavelet power. (E) Cohort effect hypotheses for a three year-old based on parameter values $\{\beta_i^j\}$ that produced the population trajectory in (A). Baselines are given by dashed lines and realized rates with three different birth conditions are given by solid colored lines. Parameter values that created these curves are listed in the table inset. (F) Relationship of fecundity (left, red trend line) and survival (right, blue trend line) to density of the previous year. Density in a given year was calculated as $A(t)/K(t)$, the abundance of adult females in a given year divided by the time-dependent carrying capacity in a given year.

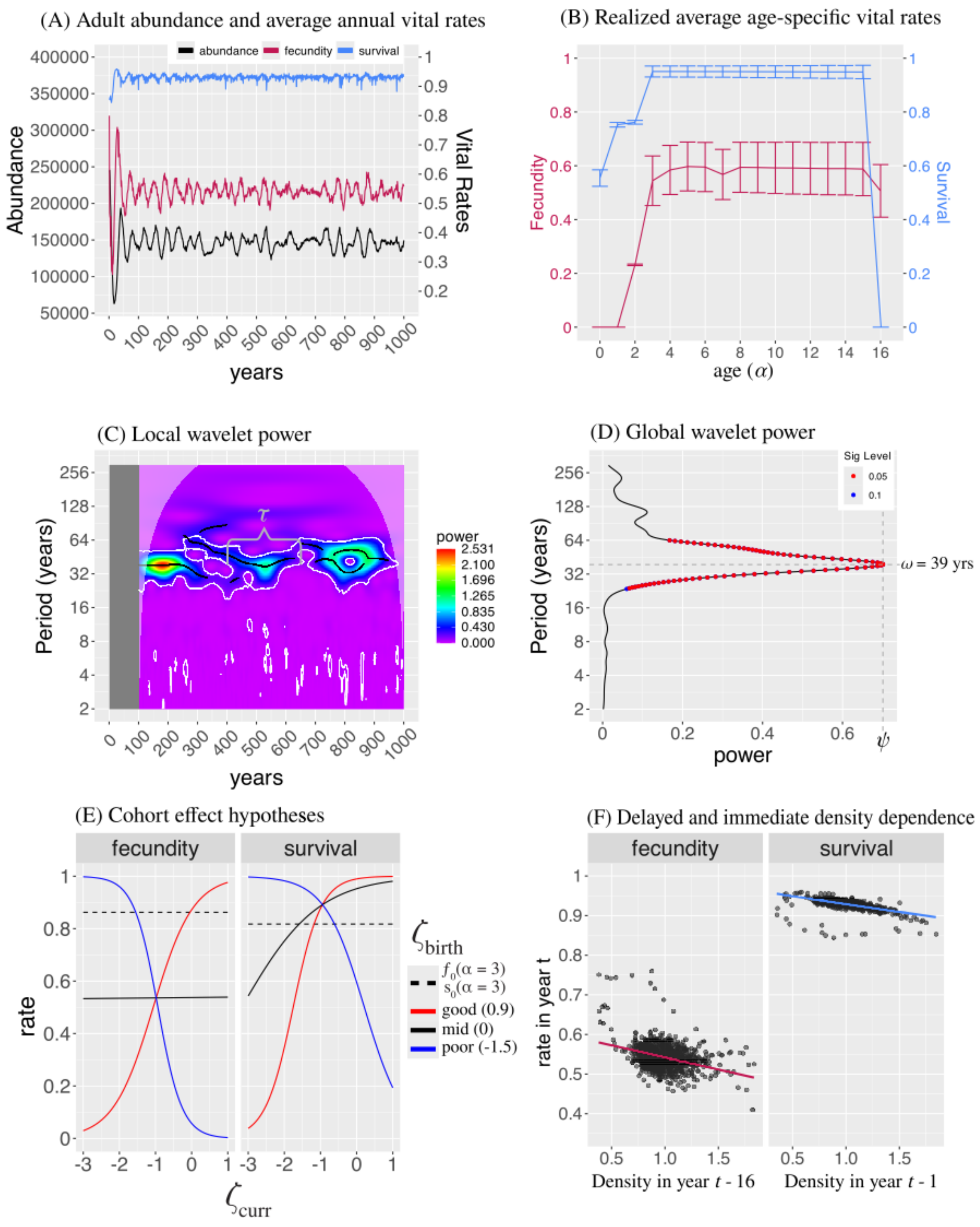

Figure C-5: Abundance, demographic rates, wavelet analysis outputs and hypotheses of cohort effects and density dependence from a sample run of the Summer AO full density dependence EMA model. (A) Adult female abundance (left) and average adult vital rates (right). Although the abundance fluctuations are relatively regular (that is, not noisy and of a fairly consistent period), they are quite small in absolute amplitude (B) Average (±SD) age-specific fecundity (left y-axis) and survival (right y-axis) across all 1,000 years of simulation. (C) Local wavelet power spectrum with 90% significance threshold outlined (white) and ridge (black) indicating local peaks

in power. The first 100 years (greyed-out portion) were trimmed prior to wavelet analysis. The COI (cone of influence, (Cazelles et al. 2008), light pink-shaded boundary region) was excluded from interpretation. Longevity ($\tau$, grey brackets) of the dominant period is marked. The local spectrum reveals the strength of the periodicity yet simultaneously highlights its episodic and disjointed nature. (D) Global average wavelet power with $\psi$ (peak power) and $\omega$ (period) marked. This $\psi$ value is on the larger end of moderate power despite the discontinuities revealed by the local spectrum. (E) Cohort effect hypotheses for a three year-old based on parameter values $\{\beta_i^j\}$ that produced the population trajectory in (A). Baselines are given by dashed lines and realized rates with three different birth conditions are given by solid colored lines. Parameter values that created these curves are listed in the table inset. (F) Relationship of fecundity (left, red trend line) to 16-year lagged density and survival (right, blue trend line) to density of the previous year (both fecundity and survival were functions of both the 16-year lagged and previous year density, however only two of these four relationships are shown). Density in a given year was calculated as $A(t)/K(t)$, the abundance of adult females in a given year divided by the time-dependent carrying capacity in a given year.

### C.4.2 No meaningful periodicity

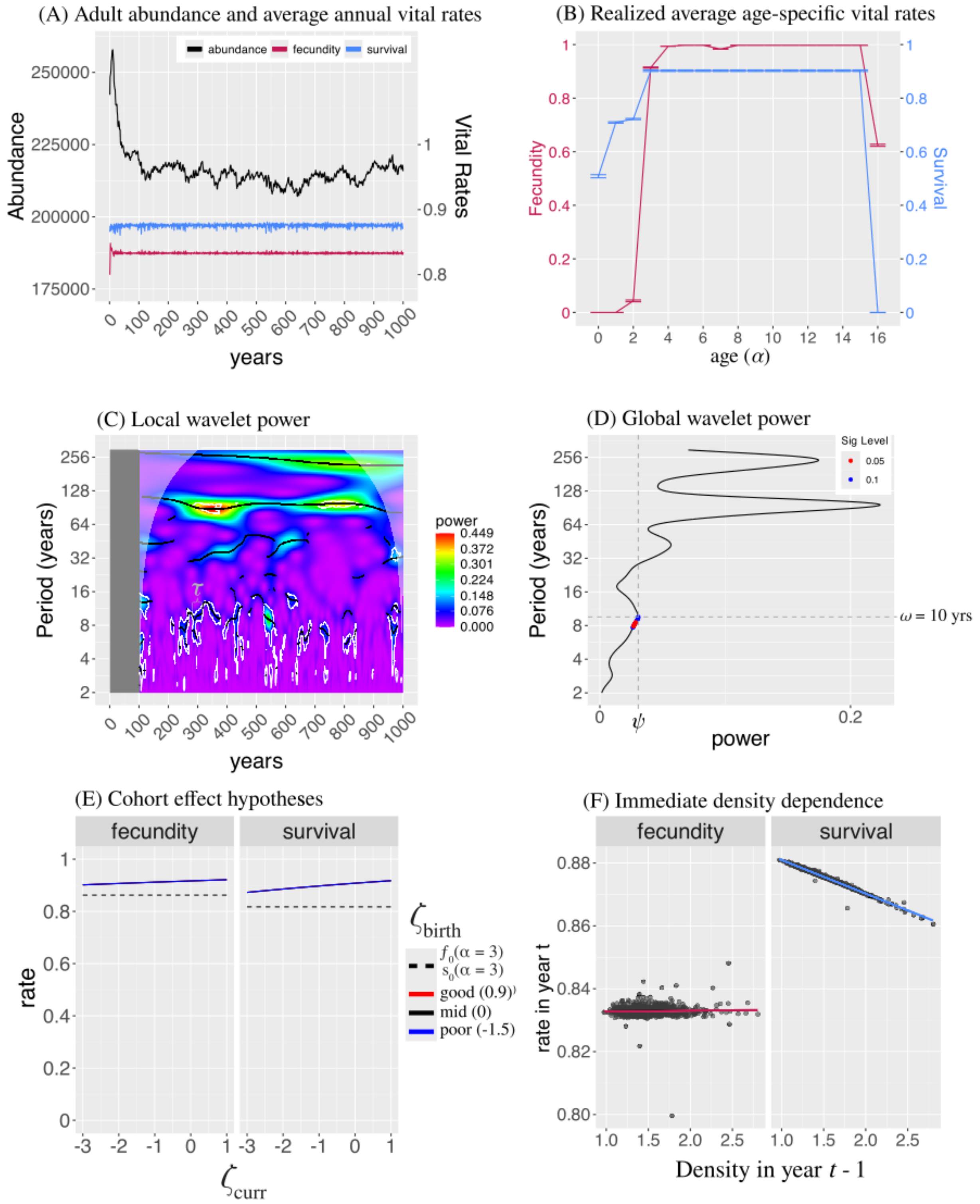

Figure C-6: Abundance, demographic rates, wavelet analysis outputs and hypotheses of cohort effects and density dependence from a sample run of the Winter freezing rain immediate density dependence CCA model. (A) Adult female abundance (left) and average adult vital rates (right). While abundance does fluctuate, it does so noisily and with minimal magnitude. Though not at a constant equilibrium, the population is also not meaningfully periodic. (B) Average (±SD) age-specific fecundity (left y-axis) and survival (right

y-axis) across all 1,000 years of simulation. (C) Local wavelet power spectrum with 90% significance threshold outlined (white) and ridge (black) indicating local peaks in power. The first 100 years (greyed-out portion) were trimmed prior to wavelet analysis. The COI (cone of influence, (Cazelles et al. 2008), light pink-shaded boundary region) was excluded from interpretation. Longevity ($\tau$, grey brackets) of the dominant period is marked. (D) Global average wavelet power with $\psi$ (peak power) and $\omega$ (period) marked. The abundance timeseries does technically have a period with statistically significant global wavelet power, however the power is negligible. (E) Cohort effect hypotheses for a three year-old based on parameter values $\{\beta_i^j\}$ that produced the population trajectory in (A). Baselines are given by dashed lines and realized rates with three different birth conditions are given by solid colored lines. Parameter values that created these curves are listed in the table inset. (F) Relationship of fecundity (left, red trend line) and survival (right, blue trend line) to density of the previous year. Density in a given year was calculated as $A(t)/K(t)$, the abundance of adult females in a given year divided by the time-dependent carrying capacity in a given year.

### C.4.3 Periodicity induced by weak drivers

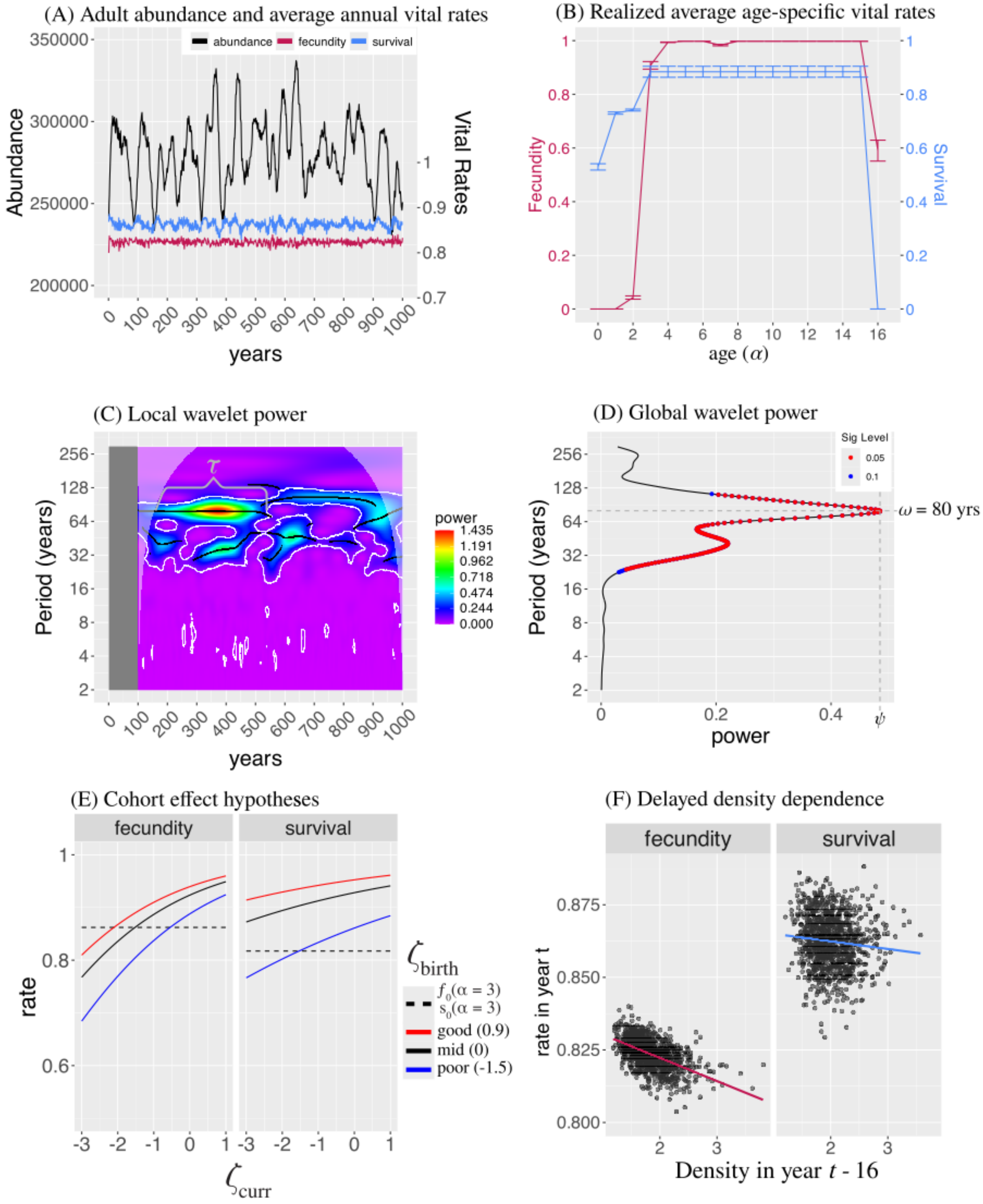

Figure C-7: Abundance, demographic rates, wavelet analysis outputs and hypotheses of cohort effects and density dependence from a sample run of the March 31st snow delayed density dependence SSA model. (A) Adult female abundance (left) and average adult vital rates (right). (B) Average (±SD) age-specific fecundity (left y-axis) and survival (right y-axis) across all 1,000 years of simulation. (C) Local wavelet power spectrum with 90% significance threshold outlined (white) and ridge (black) indicating local peaks in power. The first 100 years (greyed-out portion) were trimmed prior to wavelet analysis. The COI (cone of influence, (Cazelles et al. 2008), light

pink-shaded boundary region) was excluded from interpretation. Longevity ($\tau$, grey brackets) of the dominant period is marked. (D) Global average wavelet power with $\psi$ (peak power) and $\omega$ (period) marked. (E) Cohort effect hypotheses for a three year-old based on parameter values $\{\beta_i^j\}$ that produced the population trajectory in (A). Baselines are given by dashed lines and realized rates with three different birth conditions are given by solid colored lines. Parameter values that created these curves are listed in the table inset. The values of $\beta_1$ for both fecundity and survival are small, especially compared to the other parameter values, indicating a weaker cohort effect. (F) Relationship of fecundity (left, red trend line) and survival (right, blue trend line) to 16-year lagged density. Density in a given year was calculated as $A(t)/K(t)$, the abundance of adult females in a given year divided by the time-dependent carrying capacity in a given year. Both vital rates show only weak relationships to lagged density.

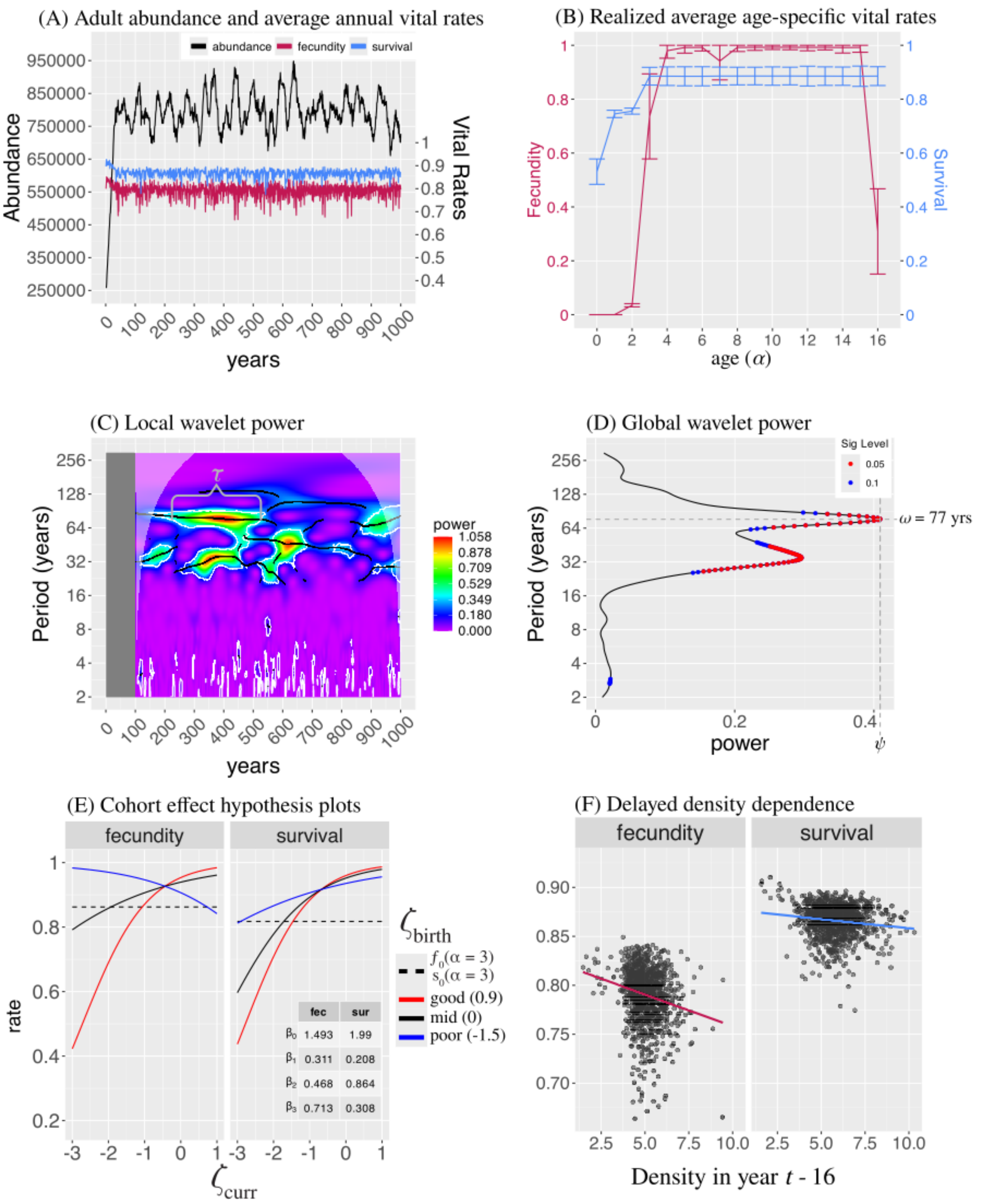

Figure C-8: Abundance, demographic rates, wavelet analysis outputs and hypotheses of cohort effects and density dependence from a sample run of the June 20 GDD delayed density dependence EMA model. (A) Adult female abundance (left) and average adult vital rates (right). (B) Average (±SD) age-specific fecundity (left y-axis) and survival (right y-axis) across all 1,000 years of simulation. (C) Local wavelet power spectrum with 90% significance threshold outlined (white) and ridge (black) indicating local peaks in power. The first 100 years (greyed-out portion) were trimmed prior to wavelet analysis. The COI (cone of influence, (Cazelles et al. 2008), light pink-shaded boundary region) was excluded from interpretation. Longevity ($\tau$, grey brackets) of the dominant period is marked. (D)

Global average wavelet power with $\psi$ (peak power) and $\omega$ (period) marked. (E) Cohort effect hypotheses for a three year-old based on parameter values $\{\beta_i^j\}$ that produced the population trajectory in (A). Baselines are given by dashed lines and realized rates with three different birth conditions are given by solid colored lines. Parameter values that created these curves are listed in the table inset. The values of $\beta_1$ ($\zeta_{birth}$ coefficient) for both fecundity and survival are small and the value of $\beta_3$ (interaction of $\zeta_{birth}$ and $\zeta_{\text{curr}}$ coefficient) for survival is similarly small. This results in a weaker cohort effect for survival, but an intermediate cohort effect for fecundity. (F) Relationship of fecundity (left, red trend line) and survival (right, blue trend line) to 16-year lagged density. Density in a given year was calculated as $A(t)/K(t)$, the abundance of adult females in a given year divided by the time-dependent carrying capacity in a given year. Both vital rates have negative relationships to lagged density, however the trends are not steep enough to constitute strongly overcompensatory density dependence.

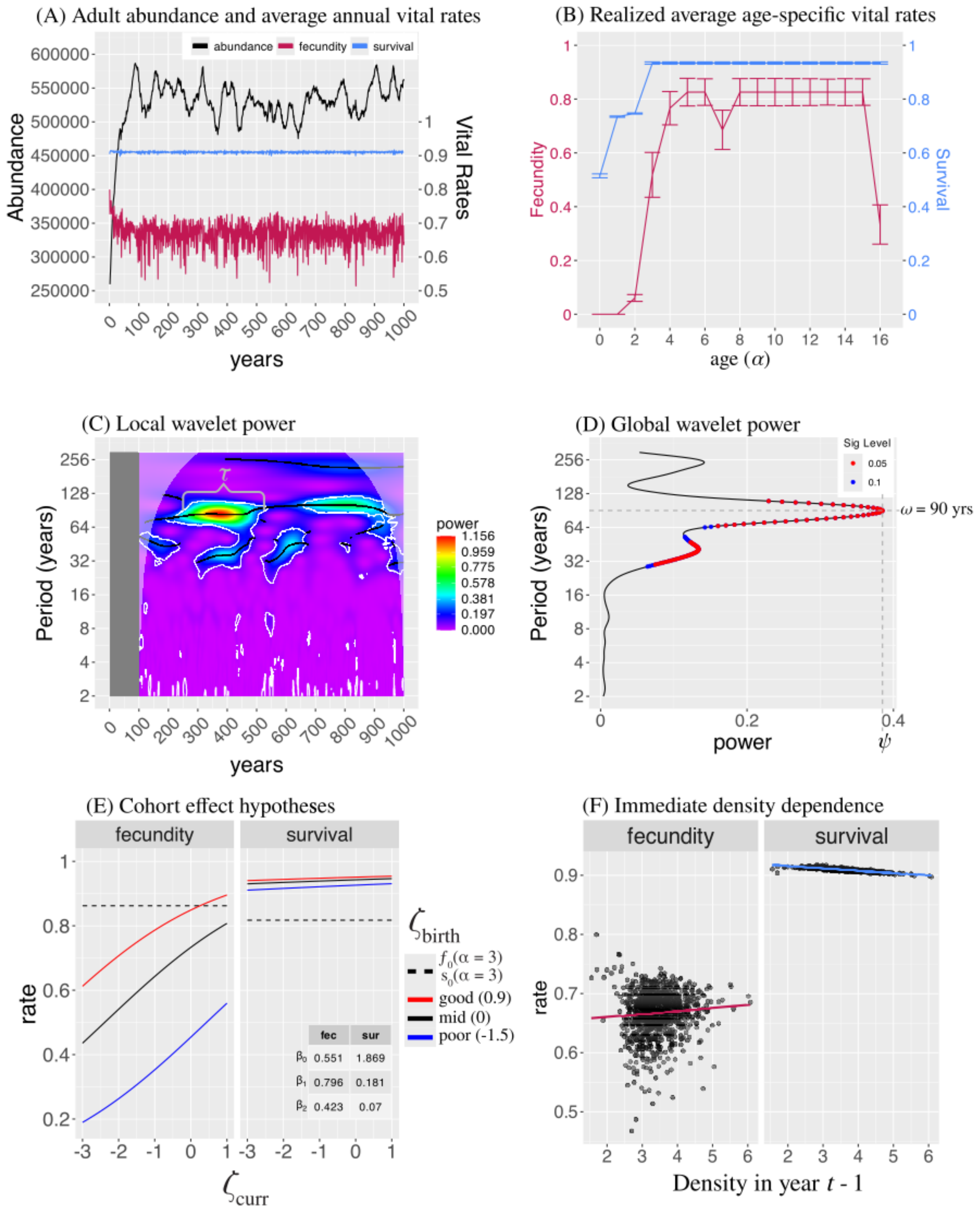

| | fec | sur |
|---|---|---|
| $\beta_0$ | 0.551 | 1.869 |
| $\beta_1$ | 0.796 | 0.181 |
| $\beta_2$ | 0.423 | 0.07 |

Figure C-9: Abundance, demographic rates, wavelet analysis outputs and hypotheses of cohort effects and density dependence from a sample run of the Summer OI immediate density dependence SSA model. (A) Adult female abundance (left) and average adult vital rates (right). (B) Average (±SD) age-specific fecundity (left y-axis) and survival (right y-axis) across all 1,000 years of simulation. (C) Local wavelet power spectrum with 90% significance threshold outlined (white) and ridge (black) indicating local peaks in power. The first 100 years (greyed-out portion) were trimmed prior to wavelet analysis. The COI (cone of influence, (Cazelles et al. 2008), light pink-shaded boundary region) was excluded from interpretation. Longevity ($\tau$, grey brackets) of the dominant period is marked. (D) Global average wavelet power with $\psi$ (peak power) and $\omega$ (period) marked. (E) Cohort effect hypotheses for a three year-old based

on parameter values $\{\beta_i^j\}$ that produced the population trajectory in (A). Baselines are given by dashed lines and realized rates with three different birth conditions are given by solid colored lines. Parameter values that created these curves are listed in the table inset. Though the $\beta_1^f$ (fecundity $\zeta_{birth}$ coefficient) value is high, resulting in a large cohort effect for fecundity, the $\beta_1^s$ (survival $\zeta_{birth}$ coefficient) value is low, resulting in only a minor cohort effect for survival. (F) Relationship of fecundity (left, red trend line) and survival (right, blue trend line) to density of the previous year. Density in a given year was calculated as $A(t)/K(t)$, the abundance of adult females in a given year divided by the time-dependent carrying capacity in a given year. Fecundity, the vital rate with the larger cohort effect, shows a counter-intuitively inverse trend regarding density dependence, while survival, the vital rate with the smaller cohort effect, shows almost no dependence on density.

## C.5 Round 2 detailed results

| Scenario | $\sum_i S_i$ for $V(\psi)$ | $\sum_i S_i$ for $V(\tau)$ | $\rho$ | $\Psi$ | $\bar{\psi}$ | $\Omega$ | $\bar{\omega}$ | T | $\bar{\tau}$ |
|---|---|---|---|---|---|---|---|---|---|
| Winter AO | 0.460 | 0.034 | 0.780 | 1.18 | 0.36 | 30.7 | 90.9 | 634 | 293 |
| Summer AO | 0.526 | 0.167 | 0.777 | 1.34 | 0.34 | 28.2 | 60.4 | 818 | 257 |
| Total AO | 0.499 | 0.195 | 0.772 | 1.40 | 0.40 | 52.7 | 54.2 | 828 | 358 |
| Summer OI | 0.532 | 0.282 | 0.769 | 1.30 | 0.36 | 32.4 | 50.3 | 766 | 185 |
| June 20 GDD | 0.448 | 0.097 | 0.774 | 1.20 | 0.38 | 32.0 | 49.2 | 599 | 226 |
| March 31 snow | 0.441 | 0.126 | 0.778 | 1.23 | 0.39 | 32.0 | 49.9 | 595 | 249 |
| Winter freezing rain | 0.376 | 0.068 | 0.764 | 1.27 | 0.31 | 32.0 | 56.2 | 755 | 180 |

Table C-2: Model outputs from the second round of simulation, organized by environmental scenario. The sum of first-order indices ($\sum_i S_i$) gives the proportion of variance attributable solely to first-order effects (Puy et al. 2022). The remaining metrics are explained in Table C-1.. Differences in maximal power ($\Psi$) are evident, however the scenarios were roughly equivalent regarding the average power ($\bar{\psi}$). Longevity was more variable, both in maximal ($T$) and average ($\bar{\tau}$) values. All scenarios are capable of yielding strong and lasting periodicity but vary in how influential higher-order interactions are to periodicity.

| Metric | Average value | Highest value | Lowest value |
|---|---|---|---|
| $\sum_i S_i$ | 0.468 | 0.532 | 0.376 |
| S: $\beta_0^f$ | 0.004 | 0.011 | 0.000 |
| T: $\beta_0^f$ | 0.047 | 0.061 | 0.037 |
| S: $\beta_1^f$ | 0.001 | 0.004 | 0.000 |
| T: $\beta_1^f$ | 0.020 | 0.024 | 0.016 |
| S: $\beta_2^f$ | 0.000 | 0.000 | 0.000 |
| T: $\beta_2^f$ | 0.020 | 0.030 | 0.013 |
| S: $\beta_0^s$ | 0.361 | 0.401 | 0.324 |
| T: $\beta_0^s$ | 0.829 | 0.870 | 0.800 |
| S: $\beta_1^s$ | 0.055 | 0.086 | 0.017 |
| T: $\beta_1^s$ | 0.275 | 0.351 | 0.229 |
| S: $\beta_2^s$ | 0.047 | 0.059 | 0.024 |
| T: $\beta_2^s$ | 0.512 | 0.589 | 0.453 |

Table C-3: Sobol’ indices for output metric $\psi$ summarized across the seven empirically motivated scenarios. The first-order index (S) quantifies the influence of a single parameter on the variance of $\tau$ and the total-order index (T) quantifies the influence of that parameter individually along with all of its interactions with other parameters. The sum of first-order indices ($\sum_i S_i$) gives the proportion of variance attributable solely to first-order effects (Puy et al. 2022). Indices, both first-order and total-order, for fecundity terms are notably reduced compared to those for survival terms.

| Metric | Average value | Highest value | Lowest value |
|---|---|---|---|
| $\sum_i S_i$ | 0.164 | 0.346 | 0.034 |
| S: $\beta_0^f$ | 0.013 | 0.027 | 0.000 |
| T: $\beta_0^f$ | 0.126 | 0.167 | 0.109 |
| S: $\beta_1^f$ | 0.003 | 0.012 | 0.000 |
| T: $\beta_1^f$ | 0.053 | 0.091 | 0.038 |
| S: $\beta_2^f$ | 0.001 | 0.010 | 0.000 |
| T: $\beta_2^f$ | 0.076 | 0.104 | 0.054 |
| S: $\beta_0^s$ | 0.096 | 0.216 | 0.028 |
| T: $\beta_0^s$ | 0.916 | 0.962 | 0.881 |
| S: $\beta_1^s$ | 0.001 | 0.007 | 0.000 |
| T: $\beta_1^s$ | 0.363 | 0.470 | 0.233 |
| S: $\beta_2^s$ | 0.024 | 0.043 | 0.000 |
| T: $\beta_2^s$ | 0.851 | 0.927 | 0.747 |

Table C-4: Sobol' indices for output metric $\tau$ summarized across the seven empirically motivated scenarios. The first-order index (S) quantifies the influence of a single parameter on the variance of $\tau$ and the total-order index (T) quantifies the influence of that parameter individually along with all of its interactions with other parameters. The sum of first-order indices ($\sum_i S_i$) gives the proportion of variance attributable solely to first-order effects (Puy et al. 2022). Indices, both first-order and total-order, for fecundity terms are notably reduced compared to those for survival terms.

# Appendix D: Basic properties of vital rate functions

In order to be mathematically and biologically valid, the function governing the vital rates ought to satisfy the following basic properties:

1. When density tends to infinity, the vital rate in question should shrink to zero,
2. Without Allee effects, when density tends to zero (or one), the vital rate should tend to its maximum possible value,
3. The density-dependent, environment-dependent model should tend to the solely density-dependent model whenever environmental variance tends to zero.

Note that the density-dependent, environment-dependent model does not tend to the solely environment-dependent model without either explicitly changing the equations or specifying that density tends toward a value dependent on the environmental value (specifically density tends toward $2K_0\left(\frac{e^{\varepsilon(t)}}{1+e^{\varepsilon(t)}}\right)(1-\varepsilon(t))$. This is because our framework has the environment operating *through* carrying capacity for all density dependent models (e.g. Roughgarden 1975). One cannot remove the influence of density without also removing the influence of the environment without altering the equations.

It suffices to prove the function $G(\alpha, t)$ for the CCA hypothesis given in the first row of Table A-1 (Table 2 in the main text) satisfies these properties. Trivially, it follows that $G(\alpha, t)$ for the other hypotheses, and that the vital rate functions $f$ and $s$, given in equation 7, also satisfy these properties.

Let $l(\alpha, t) = \beta_0 \operatorname{logit}\big(f_0(\alpha)\big) + \beta_2 \zeta(t)$.

1. $\lim_{A(t)\to\infty} \zeta(t) = \lim_{A(t)\to\infty} 1 - \frac{A(t)}{K(t)} = -\infty$

   $\lim_{\zeta\to-\infty} l(\alpha, t) = \lim_{\zeta\to-\infty} \beta_0 \operatorname{logit}\big(f_0(\alpha)\big) + \beta_2\zeta(t) = \beta_0 \operatorname{logit}\big(f_0(\alpha)\big) - \infty = -\infty$

   $$\lim_{l(\alpha,t)\to-\infty} G(\alpha, t) = \lim_{l(\alpha,t)\to-\infty} \frac{e^{l(\alpha,t)}}{1 + e^{l(\alpha,t)}} = \frac{0}{1+0} = 0$$

   Therefore, $\lim_{A(t)\to\infty} G(\alpha, t) = 0.$ ■

2. Note that, for a specific age-baseline fecundity $f_0(\alpha)$ and a specific set of parameters $\{\beta_0, \beta_2\}$, given that $\max_{-1\le\varepsilon(t)\le1} \zeta = 1$, the maximum possible realized fecundity occurs when $G(\alpha, t) = \frac{e^{\beta_0 \operatorname{logit}(f_0(\alpha))+\beta_2}}{1+e^{\beta_0 \operatorname{logit}(f_0(\alpha))+\beta_2}}$.

   $$\lim_{A(t)\to0} \zeta(t) = \lim_{A(t)\to0} 1 - \frac{A(t)}{K(t)} = 1 - \frac{0}{K(t)} = 1$$

   $\lim_{\zeta\to1} l(\alpha, t) = \lim_{\zeta\to1} \beta_0 \operatorname{logit}\big(f_0(\alpha)\big) + \beta_2\zeta(t) = \beta_0 \operatorname{logit}\big(f_0(\alpha)\big) + \beta_2$

   $$\lim_{l(\alpha,t)\to\beta_0 \operatorname{logit}(f_0(\alpha))+\beta_2} G(\alpha, t) = \frac{e^{\beta_0 \operatorname{logit}(f_0(\alpha))+\beta_2}}{1 + e^{\beta_0 \operatorname{logit}(f_0(\alpha))+\beta_2}}$$

   Therefore, $\lim_{A(t)\to0} G(\alpha, t) = \frac{e^{\beta_0 \operatorname{logit}(f_0(\alpha))+\beta_2}}{1+e^{\beta_0 \operatorname{logit}(f_0(\alpha))+\beta_2}}.$ ■

3. Note that the model with a fixed carrying capacity $K_0$ would be the solely density-dependent model without environmental variation, i.e. when $\varepsilon(t) = 0$. In that case, we have that

   $$\lim_{\varepsilon(t)\to0} K(t) = \lim_{\varepsilon(t)\to0} 2K_0\left(\frac{e^{\varepsilon(t)}}{1 + e^{\varepsilon(t)}}\right) = 2K_0\left(\frac{e^0}{1+e^0}\right) = 2K_0\left(\frac{1}{2}\right) = K_0.$$ ■

   If the environmental variation is 0 but for some reason the actual constant environmental value is nonzero, i.e. when $\varepsilon(t) = c$ for some constant $c$, we can simply recenter a new constant carrying capacity $K_0^*$ such that $K_0^* = 2K_0\left(\frac{e^c}{1+e^c}\right)$. Then we have that $K(t) = K_0^* \; \forall \, t.$ ■

# Appendix E: Confirmation of Sobol' Sensitivity Analysis Results via Linear Discriminant Analysis

Results from the Sobol' sensitivity analysis were additionally confirmed via Linear Discriminant Analysis (LDA), a dimension reduction technique similar to Principal Component Analysis (PCA). While PCA is focused on identifying components that capture the most variance in the parameters, LDA is focused on identifying components that maximally separate different classes of parameters in order to explain variance in model output (Zhao et al. 2024). To implement this, the original data are labelled according to model class and then the trace of the between-class scatter matrix is maximized while the trace of the within-class scatter matrix is minimized. The data were the parameter sets $\{\beta_i^j\}$, and three classes related to power were used: runs resulting in $\psi$ in the 10$^{th}$ percentile for the model, runs resulting in $\psi$ in the bottom 90$^{th}$ percentile, and runs resulting in no periodicity at all. The recent primer by Zhou et al. (2024) offers a thorough review of LDA and its underpinnings, methods, and applications. LDA was performed in R 4.3.1 with the *MASS* package version 7.3-60 (Venables and Ripley 2002).

Across all seven scenarios, the results of the LDA were strikingly similar, indicating the existence of a consistent region of parameter space likely to yield strong periodicity across diverse environmental scenarios. $\beta_0^s$ was confirmed as the most significant coefficient, followed by $\beta_1^s$ and $\beta_2^s$, corroborating the findings of the Sobol' sensitivity analysis.

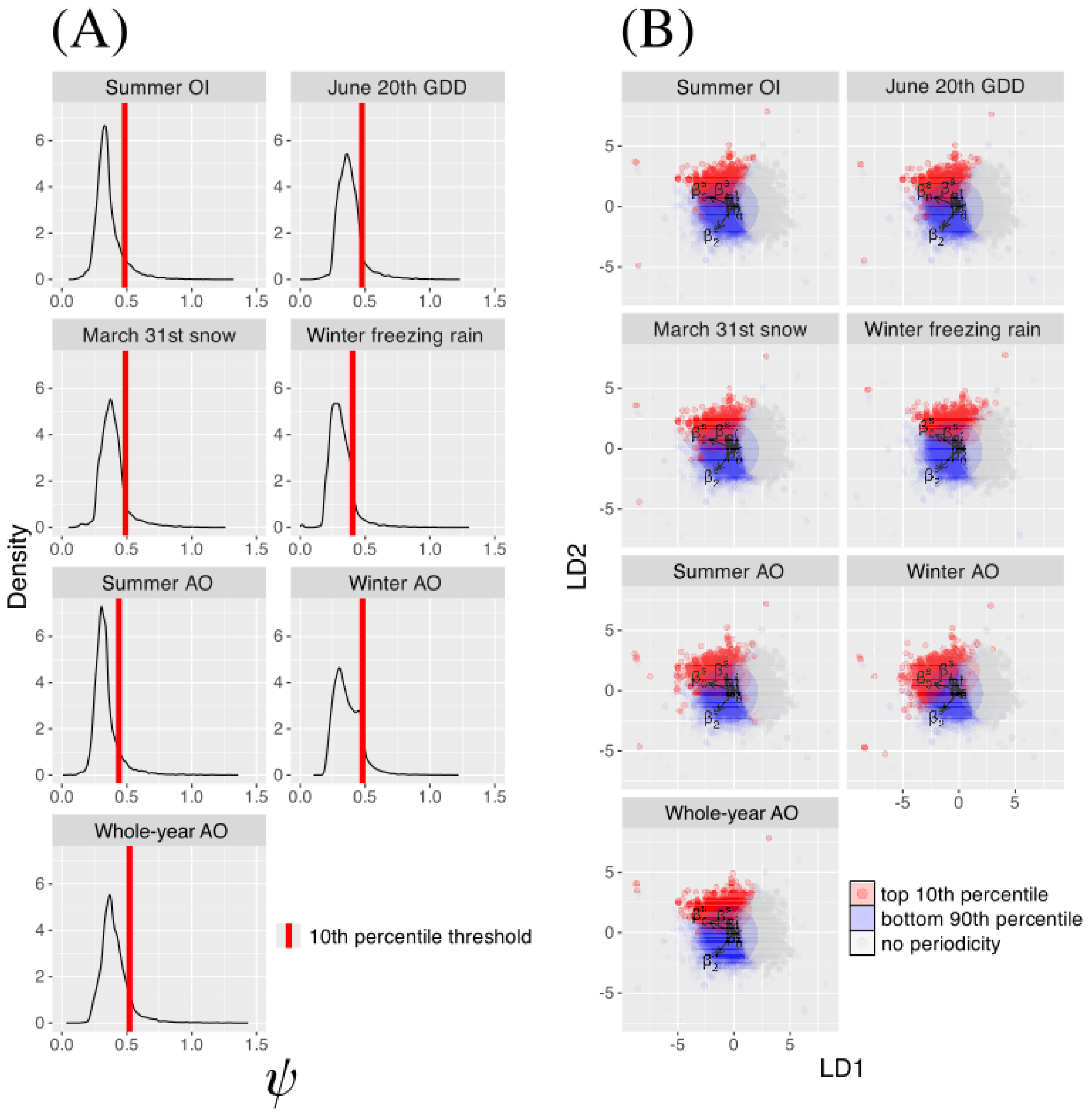

Figure E-1: Outputs from round 2 with only full density dependence SSA models (A) Distributions of $\psi$ (strength of periodicity) across periodic runs We calculated $p_{avg}$ sets by averaging $\{\beta_i^j\}$ values of runs above the 10th percentile threshold (red). (B) Linear Discriminant Analysis plots. Groups stratify runs by $\psi$: periodic runs within the 10th percentile (red), periodic runs within the bottom 90th percentile (blue), and runs without periodicity (grey). Individual runs (points) are colored by group with corresponding ellipses, containing 95% of group points, centered on group means. LDA further confirms that survival terms are the primary drivers of periodicity.

# Appendix F: Model results contrasted with empirical estimates for the Bathurst herd

Amplitude of oscillations in abundance varied widely across simulated population trajectories identified as periodic. Most models showed only moderate fluctuations in abundance; for example, the $p_{avg}$ run of the winter freezing rain scenario with the largest fluctuations has a maximum amplitude of only ~ 30,000 adult females despite abundances ~200,000. Some models showed massive gains and losses in abundance, such as the one displayed in Figure F-1 A; this run resulted in a decrease of ~220,000 adult females in just 13 years as part of a 30-year cycle with an average amplitude of nearly 100,000 individuals. Compared to the loss of 200,000+ adult females in the Bathurst herd (Adamczewski et al. 2021), the initial decline of this simulated trajectory appears similar (Figure F-1 E1). However, major discrepancies between the two time series are also apparent: the simulated trajectory rapidly recovers and begins an increasing phase right at the point where the Bathurst herd population crash accelerates. The nadir of the simulated abundance exceeds 100,000 adult females, well above the Bathurst nadir of <5000 adult females (Adamczewski et al. 2021). Further still, empirical fecundity and survival estimates differ substantially from the averages observed in the model with high amplitude, long-period fluctuations (Figure F-1 E2). For example, during the declining phase of the simulation, the overall herd fecundity was below the 95% CI of empirical estimates of herd fecundity. Survival estimates are not available during the early years of the Bathurst decline; however, later estimates of Bathurst survival are markedly lower than simulated values. The accelerated decline and pronounced lack of recovery in the Bathurst herd represents a stark departure from even the most exaggerated declining phases of simulated population cycles.

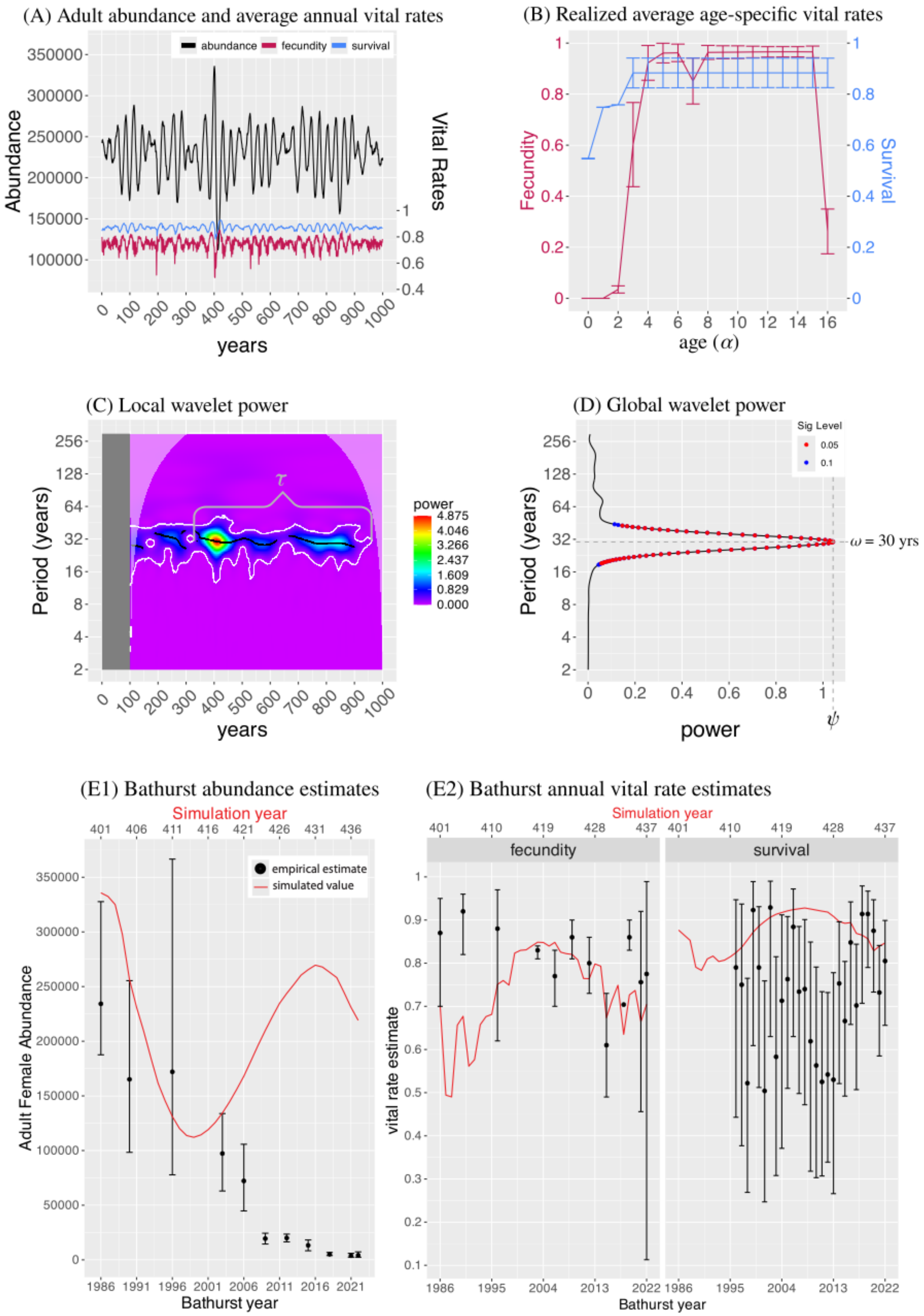
(A) Adult abundance and average annual vital rates
abundance
fecundity
survival
Abundance
Vital Rates
years
(B) Realized average age-specific vital rates
Fecundity
Survival
age (α)
(C) Local wavelet power
τ
power
4.875
4.046
3.266
2.437
1.609
0.829
0.000
Period (years)
years
(D) Global wavelet power
Sig Level
0.05
0.1
ω = 30 yrs
Period (years)
power
ψ
(E1) Bathurst abundance estimates
Simulation year
empirical estimate
simulated value
Adult Female Abundance
Bathurst year
(E2) Bathurst annual vital rate estimates
Simulation year
fecundity
survival
vital rate estimate
Bathurst year

Figure F-1: Abundance, demographic rates, wavelet analysis outputs and comparison with Bathurst decline from the Winter AO full density dependence SSA model run with $p_{top}$ parameters. (A) Adult female abundance (left) and annual average adult female vital rates (right). Abundance cycled throughout, but a spike around year 400 more than doubled the usual amplitude before stabilizing. This extreme cycle was not clearly reflected in average vital rates, suggesting cohort effects were masked in aggregated rates. (B) Age-specific fecundity (left) and survival (right) with standard deviation bars. Fecundity was higher and less variable than in Figure 4B of the main text. Survival patterns were similar to Figure 4B of the main text, though variance in calves and yearlings remained minimal, likely due to exclusion of birth effect terms to prevent conflation with current-year effects. (C) Local wavelet power spectrum with 90% significance threshold outlined (white) and ridge (black) indicating the local peaks in power. The first 100 years (greyed-out portion) were trimmed prior to wavelet analysis. The COI (cone of influence, (Cazelles et al. 2008), light pink-shaded boundary region) was excluded from interpretation. Longevity ($\tau$, grey brackets) of the dominant period is marked. The amplified cycle from 400-430 years results in a spike in local power that temporarily far exceeds the maximum local wavelet power in Figure 4C of the main text. (D) Global average wavelet power with peak power $(\psi)$ and period $(\omega)$ marked. Power is lower than in Figure 4D of the main text but remains high. (E) Simulated and empirical Bathurst estimates of (E1) adult female abundances and (E2) adult female average fecundity and survival. Black points are empirical estimates with 95% confidence intervals (Adamczewski et al. 2021) and red lines are simulation values. Simulated dynamics track Bathurst decline until ~year2000 (simulation year 415), after which only the simulation recovers. Simulated fecundity is reduced compared to empirical estimates during the early years of the Bathurst decline, but more closely align with empirical estimates in later years. Simulated survival is markedly higher than empirical survival, particularly in later years as the simulated and real trajectories further diverge. This highlights the importance of survival in facilitating the recovery of the simulated abundance trajectory.

# Appendix G: Model-adjusted age-specific fecundity baselines

The empirical data come from individuals sampled from Thomas and Barry (1990), who sampled individuals from the Beverly herd during November, December, and March from 1980-1987. They recorded the total number of sampled females, here denoted $N_T$, the number of sampled individuals of age $\alpha$, here denoted $N_T(\alpha)$, and the number of individuals of age $\alpha$ that were pregnant, here denoted $N_p(\alpha)$. The average total fecundity for the Beverly herd during these years, here denoted $f_0$, was 72%, calculated as

$$f_0 = \frac{\sum_{\alpha=2}^{16} N_p(\alpha)}{N_T}. \qquad (A1)$$

We calculated the age-specific fecundity,

$$f_0(\alpha) = \frac{N_p(\alpha)}{N_T(\alpha)}. \qquad (A2)$$

As a proxy for the age distribution of adult females in the herd, we used the proportion of sampled females of age $\alpha$,

$$q(\alpha) = \frac{N_T(\alpha)}{N_T}. \qquad (A3)$$

We then calculated the proportion of the total herd fecundity attributable to individuals of age $\alpha$,

$$P(\alpha) = q(\alpha) f_0(\alpha), \qquad (A4)$$

with $\sum_{\alpha=2}^{16} P(\alpha) = f_0 = 0.72$.

For situations with the same maximum age of 16, we used an asterisk to indicate terms used in the final model (those adjusted to match the desired total herd fecundity). We denoted the model-adjusted total herd fecundity by $f_0^*$ and calculated $P^*(\alpha)$, the model-adjusted proportion of total herd fecundity attributable to individuals of age $\alpha$,

$$P^*(\alpha) = \frac{f_0^*}{f_0} P(\alpha). \qquad (A5)$$

We used the above to calculate $f_0^*(\alpha)$, the model-adjusted age-specific fecundity,

$$f_0^*(\alpha) = \min\left(0.99, \frac{P^*(\alpha)}{q(\alpha)}\right), \qquad (A6)$$

with the minimum function to ensure that $f_0^*(\alpha)$ was in the domain of the logit function to facilitate calculation of realized age-specific fecundity according to equations 10-13 in the main text.

| $\alpha$ | $N_p(\alpha)$ | $N_T(\alpha)$ | $q(\alpha)$ | $f_0(\alpha)$ | $P(\alpha)$ | $P^*(\alpha)$ | $f_0^*(\alpha)$ |
|---|---|---|---|---|---|---|---|
| 2 | 11 | 92 | 0.162 | 0.12 | 0.0194 | 0.0233 | 0.14 |
| 3 | 86 | 120 | 0.144 | 0.72 | 0.1032 | 0.1242 | 0.86 |
| 4 | 88 | 108 | 0.128 | 0.81 | 0.1043 | 0.1255 | 0.98 |
| 5 | 72 | 83 | 0.113 | 0.87 | 0.0980 | 0.1179 | 0.99 |
| 6 | 85 | 99 | 0.098 | 0.86 | 0.0841 | 0.1012 | 0.99 |
| 7 | 55 | 69 | 0.084 | 0.80 | 0.0670 | 0.0806 | 0.96 |
| 8 | 56 | 59 | 0.071 | 0.95 | 0.0674 | 0.0811 | 0.99 |
| 9 | 48 | 52 | 0.059 | 0.92 | 0.0545 | 0.0655 | 0.99 |
| 10 | 39 | 44 | 0.047 | 0.89 | 0.0417 | 0.0501 | 0.99 |
| 11 | 17 | 20 | 0.037 | 0.85 | 0.0315 | 0.0378 | 0.99 |
| 12 | 27 | 30 | 0.027 | 0.90 | 0.0243 | 0.0292 | 0.99 |
| 13 | 11 | 12 | 0.017 | 0.92 | 0.0156 | 0.0187 | 0.99 |
| 14 | 9 | 9 | 0.01 | 0.99 | 0.0099 | 0.0119 | 0.99 |
| 15 | 2 | 2 | 0.0019 | 0.99 | 0.0019 | 0.0023 | 0.99 |
| 16 | 1 | 2 | 0.001 | 0.50 | 0.0005 | 0.0006 | 0.60 |
| Total | 607 | $N_T = 801$ | 1 | NA | $f_0 = 0.72$ | $f_0^* = 0.87$ | NA |

Table G-1: Demonstration of calculation of age-specific fecundity with maximum age of 16 adjusted to match total herd fecundity of 87% using Beverly herd data from Thomas and Barry (1990), for which the maximum age was 16 and the total herd fecundity was 72%. Here $\alpha$ refers to the age individuals would have been when giving birth the following June.

For situations with a maximum age of 12, we removed individuals of age 13 and up from the data. This preserved the age-specific fecundity, $f_0(\alpha)$, from the original data, but resulted in different values for sample size, age distribution, age-specific contribution to total herd fecundity, and total herd fecundity. We used a tilde (~) to indicate values adjusted to match the desired maximum age but still reflective of the age-specific fecundity derived from the Beverly herd. We again use an asterisk to indicate terms used in the final model (those adjusted to match both the desired maximum age *and* the desired herd total herd fecundity).

After removing the older age classes, we calculated the Beverly-adjusted total number of sampled females,

$$\widetilde{N_T} = \sum_{\alpha=2}^{12} N_T(\alpha)\,, \qquad (A7)$$

and the Beverly-adjusted total herd fecundity,

$$\widetilde{f_0} = \frac{\sum_{\alpha=2}^{12} N_p(\alpha)}{\widetilde{N_T}} = 0.75. \qquad (A8)$$

We used the above to calculate the Beverly-adjusted proportion of sampled females of age $\alpha$,

$$\widetilde{q(\alpha)} = \frac{N_T(\alpha)}{\widetilde{N_T}}, \qquad (A9)$$

which we then used to calculate the Beverly-adjusted age-specific contribution to total herd fecundity,

$$\widetilde{P(\alpha)} = \widetilde{q(\alpha)} f_0(\alpha), \qquad (A10)$$

with $\sum_{\alpha=2}^{12} \widetilde{P(\alpha)} = \widetilde{f_0} = 0.75.$

We denoted the model-adjusted total herd fecundity by $f_0^*$, and calculated $P^*(\alpha)$, the model-adjusted age-specific contribution to total herd fecundity,

$$P^*(\alpha) = \frac{f_0^*}{\widetilde{f_0}} \widetilde{P(\alpha)}. \qquad (A11)$$

Finally, we calculated $f^*(\alpha)$, the adjusted age-specific fecundity,

$$f_0^*(\alpha) = \min\left(0.99, \frac{P^*(\alpha)}{\widetilde{q(\alpha)}}\right). \qquad A(12)$$

| $\alpha$ | $N_p(\alpha)$ | $N_T(\alpha)$ | $\widetilde{q(\alpha)}$ | $f_0(\alpha)$ | $\widetilde{P(\alpha)}$ | $P^*(\alpha)$ | $f_0^*(\alpha)$ |
|---|---|---|---|---|---|---|---|
| 2 | 11 | 92 | 0.119 | 0.12 | 0.0142 | 0.0164 | 0.14 |
| 3 | 86 | 120 | 0.155 | 0.72 | 0.1108 | 0.1281 | 0.83 |
| 4 | 88 | 108 | 0.139 | 0.81 | 0.1134 | 0.1311 | 0.94 |
| 5 | 72 | 83 | 0.107 | 0.87 | 0.0928 | 0.1073 | 0.99 |
| 6 | 85 | 99 | 0.128 | 0.86 | 0.1095 | 0.1266 | 0.99 |
| 7 | 55 | 69 | 0.089 | 0.80 | 0.0709 | 0.0819 | 0.92 |
| 8 | 56 | 59 | 0.076 | 0.95 | 0.0722 | 0.0834 | 0.99 |
| 9 | 48 | 52 | 0.067 | 0.92 | 0.0619 | 0.0715 | 0.99 |
| 10 | 39 | 44 | 0.057 | 0.89 | 0.0503 | 0.0581 | 0.99 |
| 11 | 17 | 20 | 0.026 | 0.85 | 0.0219 | 0.0253 | 0.98 |
| 12 | 27 | 30 | 0.039 | 0.90 | 0.0348 | 0.0402 | 0.99 |
| Sum | 584 | $\widetilde{N_T} = 776$ | 1 | NA | $\widetilde{f_0} = 0.75$ | $f_0^* = 0.87$ | NA |

Table G-2 Demonstration of calculation of age-specific fecundity with maximum age of 12 adjusted to match total herd fecundity of 87% using Beverly herd data from Thomas and Barry (1990), for which the maximum age was 16 and the total herd fecundity was 72%. Here $\alpha$ refers to the age individuals would have been when giving birth the following June.

The resulting $f_0^*(\alpha)$ values were then used as the age-specific baseline values, $f_0(\alpha)$, in the various equations for $G_h(\cdot)$ (Eqs. 10-13 in the main text). These $G_h(\cdot)$ were then plugged into the equations for the realized age-specific fecundity, $f_h(\alpha, t, \cdot)$ (Eq. 5a in the main text).

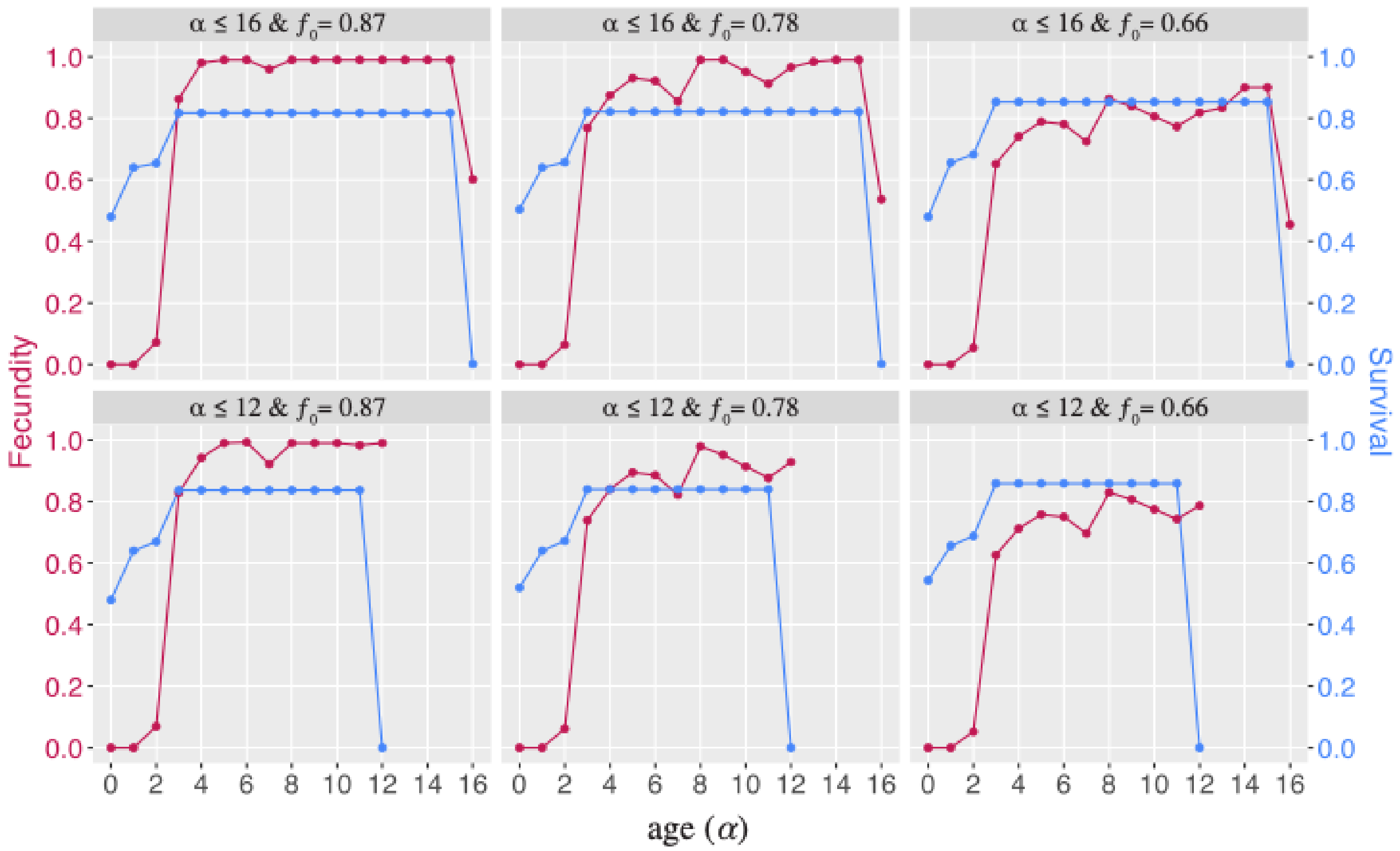

Figure G-1: Age-specific baseline fecundity (left y-axis) and survival (right y-axis) rates for situations with different total herd fecundity ($f_0$) and maximum age ($\alpha$). Models run with these values for fecundity and survival in the absence of environmental drivers, density dependence and cohort effects result in a stable age-distribution with a population multiplier of 1. The top left panel ($\alpha \leq 16$ and $f_0 = 0.87$) was the primary situation considered throughout this study. The other panels correspond to situations in the final round of simulation, designed to test full density dependence SSA model outcomes with different assumptions regarding age-structure and fecundity.

# Supplement References